\documentclass[aps,pra,reprint,superscriptaddress,showpacs,nofootinbib]{revtex4-2}
\usepackage{amsmath}
\usepackage{amsfonts}
\usepackage{amsthm}
\usepackage{amssymb}
\usepackage{verbatim}
\usepackage{graphicx}
\usepackage{enumerate}
\usepackage{lineno}
\usepackage{color,xcolor}
\usepackage[T1]{fontenc}
\usepackage{mathptmx}
\usepackage{braket}
\usepackage{todonotes}
\usepackage{soul}
\usepackage{threeparttable}
\usepackage{multirow}
\usepackage{subfigure}
\usepackage{eucal}
\usepackage{makecell}
\usepackage{diagbox}
\usepackage{accents}
\usepackage{algorithm}
\usepackage{algorithmic}
\floatname{algorithm}{Protocol}
\usepackage[bookmarks=false,colorlinks,citecolor=blue,linkcolor=blue,anchorcolor=blue,urlcolor=blue]{hyperref}

\DeclareMathAlphabet{\mathcal}{OMS}{cmsy}{m}{n}
\DeclareSymbolFont{largesymbols}{OMX}{cmex}{m}{n}

\makeatletter
\newenvironment{breakablealgorithm}
{
		\begin{center}
			\refstepcounter{algorithm}
			\hrule height.8pt depth0pt \kern2pt
			\parskip 0pt
			\renewcommand{\caption}[2][\relax]{
				{\raggedright\textbf{\fname@algorithm~\thealgorithm} ##2\par}%
				\ifx\relax##1\relax 
				\addcontentsline{loa}{algorithm}{\protect\numberline{\thealgorithm}##2}%
				\else 
				\addcontentsline{loa}{algorithm}{\protect\numberline{\thealgorithm}##1}%
				\fi
				\kern2pt\hrule\kern2pt
			}
		}
		{
		\kern2pt\hrule\relax
	\end{center}
}
\makeatother

\begin{document}
	\title{Error-mitigated inference of quantum network topology}
	\author{Jun-Hao Wei}
	\author{Xin-Yu Xu}
	\author{Shu-Ming Hu}
	\author{Nuo-Ya Yang}
	\affiliation{Hefei National Research Center for Physical Sciences at the Microscale and School of Physical Sciences, University of Science and Technology of China, Hefei 230026, China}
	\affiliation{CAS Center for Excellence in Quantum Information and Quantum Physics, University of Science and Technology of China, Hefei 230026, China}
	\author{Li Li}
	\email{eidos@ustc.edu.cn}
	\author{Nai-Le Liu}
	\email{nlliu@ustc.edu.cn}
	\author{Kai Chen}
	\email{kaichen@ustc.edu.cn}
	\affiliation{Hefei National Research Center for Physical Sciences at the Microscale and School of Physical Sciences, University of Science and Technology of China, Hefei 230026, China}
	\affiliation{CAS Center for Excellence in Quantum Information and Quantum Physics, University of Science and Technology of China, Hefei 230026, China}
	\affiliation{Hefei National Laboratory, University of Science and Technology of China, Hefei 230088, China}
	
	\date{Dec 3, 2024}
	
	\begin{abstract}
		Paramount for performances of quantum network applications are the structure and quality of distributed entanglement. Here we propose a scalable and efficient approach to reveal the topological information of unknown quantum networks, and quantify entanglement simultaneously. The scheme exploits entropic uncertainty, an operationally meaningful measure of correlation, by performing only two local measurements on each qubit. Moreover, when measurement outcomes in each node are collectively evaluated, integrating uncertainty and mutual information enables a direct count of the number of bipartite sources between any two nodes. This surpasses what is possible via applying either approach solely. Moreover, quantum error mitigation techniques including probabilistic error cancellation (PEC) and virtual distillation (VD), which have been widely applied to suppress biases in single expectation value, are further incorporated to mitigate errors in entropic quantities. We find that PEC successfully removes deviations in correlation estimations. Meanwhile, VD extends the depolarizing noise strength that allows for valid bipartite entanglement certification from 8.8\% to 26.4\%, thus substantially enhancing robustness against bias-inducing noise in practical situations. The proposal is applicable to a broad variety of platforms and helps to spur future studies toward harnessing the advantages of quantum networks.
	\end{abstract}
 
	\maketitle
	
	\section{Introduction}
	Quantum network \cite{kimble_quantum_2008,wehner_quantum_2018} has emerged as a prominent infrastructure that underpins numerous revolutionizing technologies beyond the reach of its classical counterpart, including unconditional secure communication \cite{bennett_quantum_2014,ekert_quantum_1991}, remote clock synchronization \cite{komar_quantum_2014}, distributed quantum computation \cite{jiang_distributed_2007} and distributed coherent sensing \cite{gottesman_longer-baseline_2012}. Central to the functionality of quantum networks is the generation and processing of quantum information in remote nodes. Early-stage network implementations have been demonstrated across several platforms \cite{ritter_elementary_2012,delteil_generation_2016,kalb_entanglement_2017,humphreys_deterministic_2018,stephenson_high-rate_2020,yan_entanglement_2022}, while quantum networks of increasing scale and complexity will be an irresistible trend \cite{simon_towards_2017,chen_integrated_2021,chen_implementation_2021}.
	
	Quantum entanglement \cite{horodecki_quantum_2009}, as a striking feature of quantum mechanics, plays a significant role in empowering the relevant multiuser information processing tasks. In order to understand and further harness the potential of quantum networks, it is imperative to characterize and quantify the entanglement presented in them. Growing interests are devoted to the problem of testing quantum network configurations. So far, numerous techniques have been developed leveraging network Bell inequalities \cite{chaves_polynomial_2016,rosset_nonlinear_2016,tavakoli_bell_2022,mao_certifying_2024}, covariance matrices \cite{aberg_semidefinite_2020,kraft_characterizing_2021} or fidelity constraints \cite{weinbrenner_certifying_2024}. These methods aim at certifying whether the entanglement structure of the given network is consistent with some specific network topologies, typically relying on prior knowledge of the network configuration.
	
	However, in scenarios where network providers are uncooperative or dishonest, they may only claim to distribute certain types of entangled states for application-specific purposes, while concealing or deliberately misrepresenting the topology of the underlying network. Even in a trusted scenario, physical-layer imperfections like channel noise can also disrupt the consistency of prior knowledge about the network topology, which in turn impacts subsequent network applications. Therefore, it is crucial for receiver nodes to have a method for inferring the correct network structure when prior topology information is unavailable or unreliable.
	
	Recently, entropic quantities were applied to discriminate and infer topologies of networks composed of Einstein-Podolsky-Rosen (EPR) states and multipartite Greenberger-Horne-Zeilinger (GHZ) states \cite{yang_strong_2022,chen_inferring_2023}. In Ref. \cite{yang_strong_2022}, von Neumann entropies measured at each node together with Shannon mutual information over all nodes were shown to serve as a feasible classifier of large quantum networks. To circumvent the computational complexity of multipartite Shannon mutual information, Chen \textit{et al.}\ \cite{chen_inferring_2023} suggested employing measured mutual information over two nodes instead, and showed the validity of this approach in distinguishing network topologies. Additionally, the topology of unknown quantum networks can be efficiently inferred when individual qubit measurements are available by employing either entropic quantities or covariance. Although both proposals make remarkable contributions to network characterization, the entanglement property of the distributed states remains elusive in them. Besides, noise in real experiments can lead to biased estimations of correlations and induce erroneous topology inference, which thereby raises practical issues that need to be seriously addressed.
		
	\begin{figure}[t]
		\centering
		\includegraphics[scale=0.27]{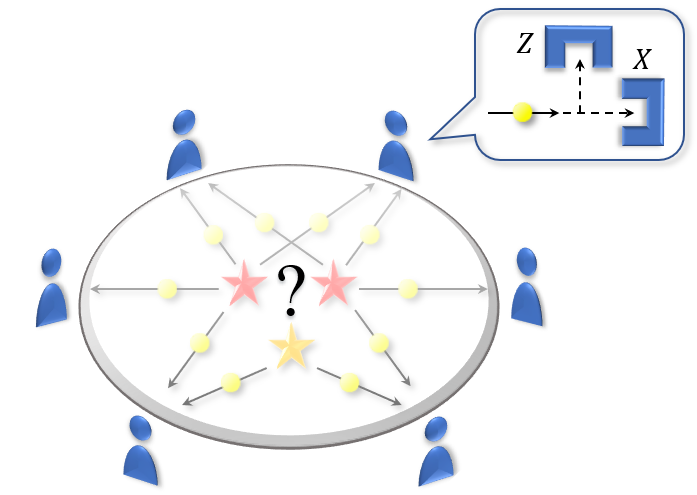}
		\caption{Schematic view on network topology inference with entropic uncertainties. The users aim to reconstruct the topology of an unknown multinode quantum network. Here, the sources, represented by colored stars, distribute entangled qubits to the users. Each qubit received from the sources is measured in two mutually unbiased bases $Z$ and $X$. The uncertainties about their outcomes are used to infer the network topology and quantitatively estimate entanglement between qubits.}
		\label{Schematic view of toppology inference}
	\end{figure}

	In this work, we propose using the entropic uncertainty to infer quantum network topology. The scenario we consider is depicted in Fig.\ \ref{Schematic view of toppology inference}. Assuming no knowledge of exact connectivity, each qubit is measured in \textit{two} mutually unbiased bases (MUBs) \cite{devetak_distillation_2005,coles_entropic_2017} and the uncertainties about their outcomes are evaluated. By means of the entropic uncertainty relations \cite{berta_uncertainty_2010}, one can completely identify the network topology meanwhile bound the entanglement between qubits from the measured uncertainties. Since entanglement is an indispensable resource for versatile quantum information applications, such as quantum teleportation \cite{bennett_teleporting_1993} and quantum key distribution \cite{ekert_quantum_1991,bennett_quantum_1992}, the proposed approach offers an easy-to-implement tool to benchmark the performance of an unknown network. As for correlations between nodes, we show that uncertainty can count the number of EPR sources when combined with measured mutual information, regardless of the network scale, thus offering more valuable information than evaluating either quantity solely. The approach possesses a quadratic scaling with respect to the number of qubits or nodes, and enables efficient and simple characterization of quantum networks.

	Moreover, we incorporate quantum error-mitigation (QEM) methods \cite{cai_quantum_2023,endo_hybrid_2021} into correlation estimations with the intention of tackling the problem of noise. Widely implemented in near-term quantum computations\ \cite{song_quantum_2019,zhang_error-mitigated_2020,urbanek_mitigating_2021,van_den_berg_probabilistic_2023,obrien_purification-based_2023}, QEM allows suppression of noise-induced errors and restoration of expectation values with decreased bias. Here, probabilistic error cancellation \cite{temme_error_2017} and virtual distillation \cite{koczor_exponential_2021,huggins_virtual_2021} are applied to achieve more accurate evaluations of statistical correlations, thereby alleviating the stringent requirement for high-fidelity entanglement in experiments. We numerically simulate the performances of the proposed scheme with and without the help of the QEM methods by optimizing measured correlations via variational quantum algorithms \cite{chen_inferring_2023}. We find that probabilistic error cancellation successfully eliminates the biases induced by dephasing noise, while virtual distillation substantially enhances depolarizing noise tolerance for certifying EPR sources. These results exhibit the augmented robustness and practicality of the error-mitigated network topology inference scheme. 
	
	\begin{figure}[b]
		\centering
		\includegraphics[scale=0.30]{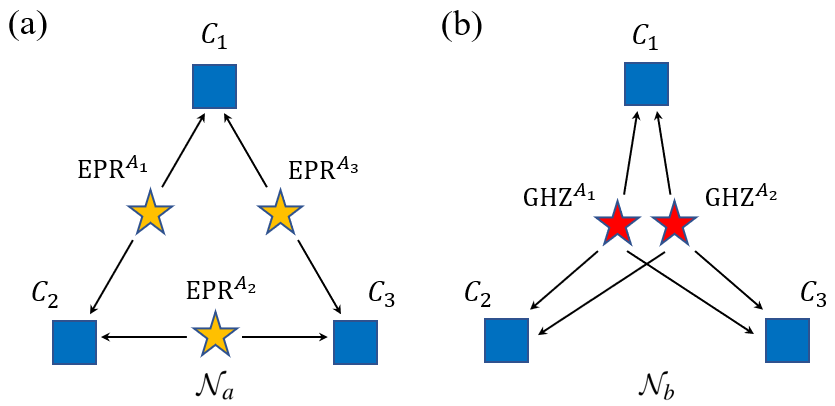}
		\caption{Two possible configurations of triangle networks. (a) Three EPR sources (orange stars) and (b) two tripartite GHZ sources (red stars) are used to connect the three nodes (blue squares) in network $\mathcal{N}_a$ and $\mathcal{N}_b$ respectively.}
		\label{Example_networks}
	\end{figure}

	We begin in Sec.\ \ref{Preliminaries} by formulating the problem of quantum network topology inference and introducing previous approaches to the problem. In Sec.\ \ref{Results}, we first overview the entropic uncertainty relations, then show that the network topology can be correctly inferred by observing the qubitwise uncertainties, whose operational implications on quantum networks are illustrated. Approaches based on different quantities are compared, while the instance where qubit measurements are unavailable is also discussed. Afterwards, Sec.\ \ref{QEM} integrates two distinct QEM methods into the network topology inference task and presents numerical tests of them by simulation. Finally, we conclude with discussions in Sec.\ \ref{Conclusion}.
	
	\section{Preliminaries on quantum network topology inference}\label{Preliminaries}
	\subsection{Problem statement}\label{Problem statement}
	Schematically, a quantum network $\mathcal{N}$ is made up of sources, labeled by $A_i,\ i = 1,2,\dots,N_A$ and nodes, labeled by $C_j,\ j = 1,2,\dots,N_C$. Each source independently distributes entangled qubits to several nodes, while each node performs local projective measurement on the received qubits. A transmitted qubit is thus represented as a link $(A_i, C_j)$ that connects a source to a node. The transmission channel is modeled by a completely positive trace-preserving map \cite{nielsen_quantum} acting on a qubit system. In general, nodes in a network may not have a shared reference frame for the encoded quantum information. Polarization encoding systems, for instance, typically require alignments that introduce non-negligible experimental overhead. Thus, the measurement bases between individual nodes can vary by arbitrary fixed but unknown qubit rotations.

	The topology of a quantum network puts limits on the extent to which each source can influence different nodes, and represents the connectivity between them. Formally speaking, two quantum networks have identical topology if and only if they have the same sources, nodes, and links, up to index permutations \cite{chen_inferring_2023}. Note that the quantum states sent by the sources in identical network topologies need not be exactly the same, as each source is characterized by the set of distributed qubits in the topological sense. Figure\ \ref{Example_networks} is an example that shows two distinct ways to connect three nodes, corresponding to different topologies of triangle networks \cite{chen_inferring_2023}. In network $\mathcal{N}_a$, all nodes are pairwise connected by three EPR sources, whereas $\mathcal{N}_b$ is a network composed of two three-qubit sources, either of which sends one qubit to each node. 
	
	The class of quantum networks under consideration is specified by the following assumptions:
	\begin{enumerate}[(A)]
		\item Each source identically and independently distributes the $m$-partite GHZ state of the form $|\text{GHZ}_m\rangle = \frac{1}{\sqrt{2}} \left( |0\rangle^{\otimes m} + |1\rangle^{\otimes m} \right) $. For $m=2$, GHZ states are just EPR states, denoted by $|\phi \rangle$.
		\item Each source sends no more than one qubit to any node.
		\item Each node performs projective measurements that are local to other nodes.
	\end{enumerate}
	While we focus on GHZ states here, we emphasize that the proposed framework admits generalizations to other entangled states, as discussed in Appendix \hyperref[Inferring the network of a W state and a generalized EPR state]{A}.
	
	The essential questions now arise: How can one infer, without \textit{a priori} information, the network topology using measurement statistics collected solely from local nodes? How can one simultaneously characterize the entanglement among the distributed qubits with complexity as low as possible? These questions are of relevance from a practical perspective, as quantum channels may be subjected to noise or sources may be prepared by a dishonest party in actual implementations. Moreover, since entanglement fuels many quantum information applications, assessing its quality with minimum effort is crucial to benchmarking the usefulness of quantum networks. The former question could be translated into determining all the links of a quantum network and was addressed in Refs. \cite{yang_strong_2022,chen_inferring_2023}. We will first review existing results in the following, while solving the latter question is the focus of this work.

	\subsection{Previous approaches}
	To determine whether two networks of EPR states and GHZ states have the same topology, Yang \textit{et al.} \cite{yang_strong_2022} used the \textit{characteristic vector} formed by von Neumann entropies of different measurement nodes
	\begin{equation}
		V_{\mathcal{N}} = \left( S(C_1)\quad S(C_2)\quad ... \quad S(C_{N_C}) \right), 
	\end{equation}
	where $S(C_j) = -\operatorname{tr}(\rho_{C_j} \log \rho_{C_j})$ is the von Neumann entropy of the reduced state $\rho_{C_j}$ on node $C_j$. It was shown that the topologies of two networks are identical if and only if their characteristic vectors are equal, conditioned on assumptions (A)-(C) along with another one: (D) each pair of nodes share no more than one source. Note that, since the one-qubit reduced state of any $m$-partite GHZ state is maximally mixed, the von Neumann entropy at each node is simply the number of sources connected to that node and its calculation demands no shared reference frame. 
	
	However, the additional assumption excludes an enormous number of networks, such as the networks in Fig.\ \ref{Example_networks}, limiting the utility of the characteristic vector in network characterization. To remove this restriction, Ref.\ \cite{yang_strong_2022} proposed employing Shannon mutual information over all nodes $I(C_1,C_2,...,C_{N_C})$, whose, however, exponential computation complexity is formidable for networks with a large number of nodes. Instead, Chen \textit{et al.}\ \cite{chen_inferring_2023} suggested evaluation of the measured mutual information between each pair of nodes
	\begin{align}
		&I(C_i, C_j)\notag \\
		=& \max_{\{ \Pi_{\vec{\alpha}_i}^{C_i} \otimes \Pi_{\vec{\alpha}_j}^{C_j} \}} \left[ H(\mathbb{P}(\vec{\alpha}_i)) + H(\mathbb{P}(\vec{\alpha}_j)) -H(\mathbb{P}(\vec{\alpha}_i, \vec{\alpha}_j))\right],
		\label{mutual information}
	\end{align}
	where $ \{ \Pi_{\vec{\alpha}_i}^{C_i} \}$ forms a complete set of projectors on all qubits in node $C_i$ with outputs $\vec{\alpha}_i$. Here $\mathbb{P}(\vec{\alpha}_i)$ is the probability distribution of $\vec{\alpha}_i$, while $H(\mathbb{P}(\vec{\alpha}_i))$ is its Shannon entropy. The measurements require optimizations so as to correctly identify the correlations. Combining with von Neumann entropy, they constructed a symmetric \textit{characteristic matrix} of a quantum network 
	\begin{equation}
		M(\mathcal{N}) = \begin{pmatrix}
			S(C_1) & I(C_1, C_2) & \cdots & I(C_1, C_{N_C})\\
			I(C_2, C_1) & S(C_2) & \cdots & I(C_2, C_{N_C})\\
			\vdots & \vdots & \ddots & \vdots\\
			I(C_{N_C}, C_1) & I(C_{N_C}, C_2) & \cdots & S(C_{N_C})
		\end{pmatrix}.
		\label{characteristic matrix}
	\end{equation}
	
	The diagonal term of $M(\mathcal{N})$ specifies the number of sources to which the node is linked, while the off-diagonal term  counts the number of sources shared between the two nodes. For example, the characteristic matrices of the networks in Fig.\ \ref{Example_networks} are
	\begin{equation}
		M(\mathcal{N}_a) = \begin{pmatrix}
			2 & 1 & 1\\
			1 & 2 & 1\\
			1 & 1 & 2
		\end{pmatrix},\ 
		M(\mathcal{N}_b) = \begin{pmatrix}
			2 & 2 & 2\\
			2 & 2 & 2\\
			2 & 2 & 2
		\end{pmatrix}.
	\end{equation}
	The defined matrix can effectively distinguish, without requiring assumption (D), the topology of two quantum networks. Notwithstanding the unique correspondence between $\mathcal{N}$ and $M(\mathcal{N})$, inferring a large network from its characteristic matrix is not straightforward. We attribute this difficulty to the ignorance of numbers of diverse sources from $M(\mathcal{N})$, as the measured mutual information cannot determine the number of qubits in each source.
	
	Furthermore, when qubitwise correlations can be evaluated, as is often the case in practice, the network topology allows efficient inference \cite{chen_inferring_2023}. This scenario amounts to regarding each qubit as an independent node, so assumption (B) can be removed. The \textit{qubitwise characteristic matrix} defined via von Neumann entropy and mutual information is
	\begin{equation}
		M^Q(\mathcal{N}) = \begin{pmatrix}
			S(Q_1) & I(Q_1, Q_2) & \cdots & I(Q_1, Q_{N_Q})\\
			I(Q_2, Q_1) & S(Q_2) & \cdots & I(Q_2, Q_{N_Q})\\
			\vdots & \vdots & \ddots & \vdots\\
			I(Q_{N_Q}, Q_1) & I(Q_{N_Q}, Q_2) & \cdots & S(Q_{N_Q})
		\end{pmatrix},
		\label{qubitwise characteristic matrix}
	\end{equation}
	where $N_Q$ is the total number of qubits. The diagonals of $M^Q(\mathcal{N})$ are all 1 as the reduced state of each qubit is maximally mixed, while the off-diagonals are either 1 or 0. Here, 1 indicates the presence of correlations between two qubits and 0 indicates no correlation. Thus, by arranging qubits from the same source together, $M^Q(\mathcal{N})$ can be transformed into a block diagonal matrix, where each block is an all-ones matrix and corresponds to a source. Hence, the number of nonzero eigenvalues of $M^Q(\mathcal{N})$ is the number of sources in $\mathcal{N}$, and each nonzero eigenvalue is the number of qubits in that source. Another viable measure of correlation between two qubits is covariance
	\begin{equation}
		\text{Cov}(Q_i, Q_j) = \max_{\{ \Pi_{{\alpha}_i}^{Q_i} \otimes \Pi_{{\alpha}_j}^{Q_j} \}} \left[\mathbb{E}({\alpha}_i, {\alpha}_j) - \mathbb{E}({\alpha}_i) \mathbb{E}({\alpha}_j) \right],
		\label{covariance}
	\end{equation}
	where $\mathbb{E}(\alpha_i)$ is the expectation of random variable $\alpha_i$. So long as the correlations between every pair of qubits are identified, the overall connectivity of network can be easily reconstructed.
	
	\section{Uncertainty-based approach for inferring the network topology}\label{Results}
	One key observation is that, though entanglement between two qubits indicates nonzero mutual information or covariance, the reverse may not hold, as the shared correlation may be classical. Particularly, for the reduced two-qubit state of multipartite GHZ state $\tau_2 = \frac{1}{2}\left( |00\rangle\langle00| + |11\rangle\langle11| \right) $, its mutual information and covariance are both 1, yet such a state is not entangled and is of no use for many quantum information tasks. In view of the significance of entanglement in quantum information science, we seek to characterize the entanglement structure of quantum networks in a more effective manner.
	\subsection{Entropic uncertainty and its applications}
	We propose here to employ the entropic uncertainty between measurement outcomes of two MUBs. To be specific, two-qubit uncertainty
	\begin{equation}
		H_{u}(Q_i | Q_j) = H(X_i | X_j) + H(Z_i | Z_j)
		\label{two-qubit uncertainty}
	\end{equation}
	is estimated, where $X_i$ and $Z_i$ are two orthogonal Pauli observables of qubit $Q_i$, and $H(X_i | X_j)$ is the conditional Shannon entropy of the outcomes of $X_i$ given those of $X_j$. The core advantage of uncertainty over either mutual information or covariance lies in its enriched operational implications by virtue of entropic uncertainty relations (EURs) \cite{coles_entropic_2017}, and its evaluation only requires two measurement settings when compared with entanglement witnesses \cite{weinbrenner_certifying_2024}.
	
	EURs impose fundamental constraints on the predictability of noncommuting observables and reveal the radical separation between classical and quantum physics. They have found wide applications in generating secure keys \cite{berta_uncertainty_2010}, detecting entanglement \cite{li_experimental_2011,prevedel_experimental_2011}, certifying randomness \cite{vallone_quantum_2014} and identifying phase transition \cite{guo_witnessing_2022}. The well-known EUR in the presence of quantum memory \cite{berta_uncertainty_2010} states that
	\begin{equation}
		H(X_i|Q_j) + H(Z_i|Q_j) \ge 1 + H(Q_i|Q_j),
		\label{EUR in presence of quantum memory}
	\end{equation}
	where $H(X_i|Q_j)$ is the conditional von Neumann entropy evaluated on the postmeasurement state (after $X_i$ is measured) given system $Q_j$, and $H(Q_i|Q_j)$ is the conditional von Neumann entropy of $Q_i$ given $Q_j$. Note that for general quantum systems $\mathcal{A}$ and $\mathcal{B}$, and for two general nondegenerate observables $R$ and $S$ with respective eigenvectors $|\mathbb{R}_r\rangle$ and $|\mathbb{S}_s\rangle$, the above EUR takes the form $H(R|\mathcal{B}) + H(S|\mathcal{B}) \geq -\log (\max_{r,s}|\langle \mathbb{R}_r|\mathbb{S}_s\rangle|^2)  + H(\mathcal{A}|\mathcal{B})$. Exploiting this EUR allows for the extension to high-dimensional quantum networks.
	
	It is known that a negative value of $H(Q_i|Q_j)$ indicates $Q_i$ and $Q_j$ are entangled. For instance, $H(Q_i|Q_j) = -1$ if the two qubits share EPR entanglement, whereas $H(Q_i|Q_j) = 0$ if they both come from a GHZ source with three or more qubits. More precisely, a lower bound on the one-way distillable entanglement $E_D$ between $Q_i$ and $Q_j$ can be obtained from their uncertainty \cite{coles_entropic_2017},
	\begin{equation}
		1 - H_{u}(Q_i | Q_j) \le -H(Q_i|Q_j) \le E_D(Q_i,Q_j),
		\label{lower bound of distillable entanglement}
	\end{equation}
	where the first inequality is from the data-processing inequality $H(X_i|Q_j)\leq H(X_i|X_j)$ \cite{lieb_proof_1973} and the second inequality is the general result of Devetak and Winter \cite{devetak_distillation_2005}. $E_D(Q_i,Q_j)$ quantifies the asymptotically optimal rate for distilling perfect EPR states from the joint state of $Q_i$ and $Q_j$, using local operations and one-way classical communication \cite{horodecki_quantum_2009}. In regard to applications in quantum cryptography, it was also shown that \cite{devetak_distillation_2005, berta_uncertainty_2010}, the amount of secret key that can be extracted from each joint state of $Q_i$ and $Q_j$ is 
	\begin{equation}
		K(Q_i,Q_j) \ge 1 - H_{u}(Q_i | Q_j).
		\label{pairwise secret key}
	\end{equation} 
	That is, two-qubit uncertainty provides a lower bound on the amount of keys that can be generated from the measured correlations of the qubits. Additionally, quantum steering \cite{uola_quantum_2020}, as another key type of quantum correlation, can be witnessed by entropic uncertainty as well. A violation of the steering inequality \cite{schneeloch_einstein-podolsky-rosen_2013}
	\begin{equation}
		H_{u}(Q_i|Q_j) \ge 1
	\end{equation}
	implies that qubit $Q_j$ can steer qubit $Q_i$, enabling one-sided device-independent information processing \cite{branciard_one-sided_2012}. Further extensions and applications of EURs are detailed in Ref. \cite{coles_entropic_2017}.
	
	\subsection{Characterizing networks from qubit measurements}
	We now show how to apply EUR to infer the entanglement structure of a quantum network and quantify the amount of entanglement simultaneously. To start, we consider the scenario where each node has access to measurements on the qubit level. In this case, each qubit effectively functions as an independent measurement node, thereby eliminating the need for assumption (B) regarding network constraints. The scenario involving collective measurements on all qubits within a local node will be discussed later in Sec. \ref{collective measurements}.
	\subsubsection{Qubitwise uncertainty matrix}
	In each node, every qubit $Q_i$, $i=1,2,...,N_Q$, is measured by two mutually unbiased qubit observables $X_i$ and $Z_i$. More precisely, denote the eigenvectors of $X_i$ and $Z_i$ as $|\mathbb{X}_{x_i}\rangle$ and $|\mathbb{Z}_{z_i}\rangle$ respectively, then $|\langle \mathbb{X}_{x_i} | \mathbb{Z}_{z_i} \rangle |^2 = 1/2,\ \forall \ x_i,z_i \in \{0, 1\}.$ Inspired by the qubitwise characteristic matrix (Eq.\ (\ref{qubitwise characteristic matrix})) defined in Ref. \cite{chen_inferring_2023}, measurement statistics of the two bases are collected to construct a \textit{qubitwise uncertainty matrix}
	\begin{align}
		&\mathbb{U}^Q(\mathcal{N}) \notag \\
		&= \begin{pmatrix}
			H_{u}(Q_1) & H_{u}(Q_1| Q_2) & \cdots & H_{u}(Q_1| Q_{N_Q})\\
			H_{u}(Q_2| Q_1) & H_{u}(Q_2) & \cdots & H_{u}(Q_2| Q_{N_Q})\\
			\vdots & \vdots & \ddots & \vdots\\
			H_{u}(Q_{N_Q}| Q_1) & H_{u}(Q_{N_Q}| Q_2) & \cdots & H_{u}(Q_{N_Q})
		\end{pmatrix},
		\label{qubitwise uncertainty matrix}
	\end{align}
	where the off-diagonal term is two-qubit uncertainty in Eq.\ (\ref{two-qubit uncertainty}) and the diagonal term is one-qubit uncertainty defined as $H_{u}(Q_i) = H(X_i) + H(Z_i)$ with $H(X_i)$ being the Shannon entropy of the outcomes of $X_i$. Note that due to the absence of shared reference frame, the measurement directions require optimizations to minimize each individual entry of uncertainty so as to identify entanglement between qubits with accuracy. We remark that if a common reference frame is pre-established across all nodes, then the measurements can be simply chosen as Pauli observables $\sigma_x$ and $\sigma_z$.
	
	We first illuminate basic properties of the defined uncertainty matrix in the noiseless circumstance. The diagonals of $\mathbb{U}^Q(\mathcal{N})$ are always 2, as each qubit is in the maximally mixed state $\mathbb{I}_2 /2$ due to the GHZ state-preparation assumption (A). Here and after, $\mathbb{I}_k$ denotes the identity matrix with rank $k$. Remarkably, taking the system $Q_j$ in Eq.\ (\ref{EUR in presence of quantum memory}) to be absent or trivial reduces the conditional entropies to unconditional ones, thus Eq.\ (\ref{EUR in presence of quantum memory}) implies $H_{u}(Q_i) \ge 1 + H(Q_i)$ with $H(Q_i)$ being the von Neumann entropy of $Q_i$ \cite{berta_uncertainty_2010}. This EUR for mixed states is tight in the considered problem, so the one-qubit uncertainty description is equivalent to the von Neumann entropy description. This equivalence persists in the case of collective measurements, as one will see in Sec. \ref{collective measurements}.

	As for the off-diagonals of $\mathbb{U}^Q(\mathcal{N})$, they offer more elaborate characterizations of the two-qubit correlations. Note that under the GHZ state-preparation assumption, there are three possible joint states for any pair of qubits: (1) If the two qubits are from one EPR source, the state will be $|\phi\rangle\langle\phi|$ with an uncertainty of 0; (2) If they are from one multipartite GHZ source, the state will be $\tau_2 = \frac{1}{2}\left( |00\rangle\langle00| + |11\rangle\langle11| \right)$ with an uncertainty of 1; (3) If they come from two different sources, the state will be $\mathbb{I}_4/4$ with an uncertainty of 2. Therefore, the uncertainties directly recognize whether two qubits originate from bipartite or multipartite entanglement source, yet the mutual information and covariance cannot, as both have value of 1.
	
\begin{figure}[tbp]
		\centering
		\includegraphics[scale=0.5]{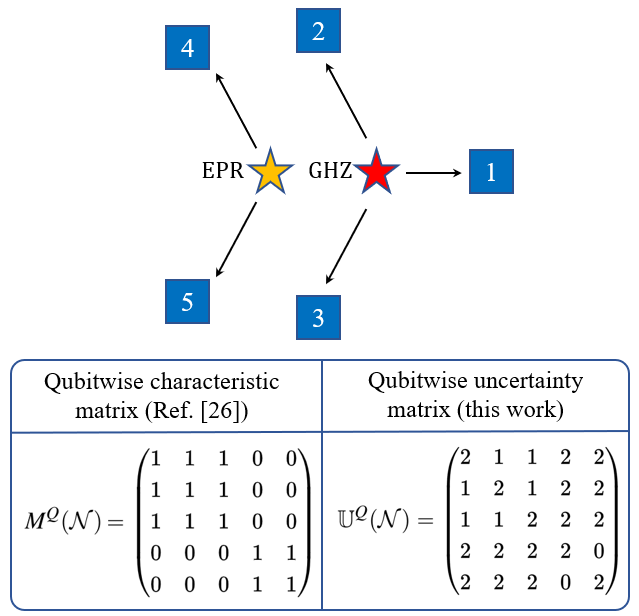}
		\caption{Example five-qubit quantum network that consists of an EPR source and a three-partite GHZ source, along with its correlation matrices defined in Eqs. (\ref{qubitwise characteristic matrix}) and (\ref{qubitwise uncertainty matrix}). Each column of either matrix stands for one qubit.}
		\label{Example_five-qubit_network}
	\end{figure}

	At this stage, one can see that the qubitwise uncertainty matrix allows for inference of the detailed network topology, as it can reduce to the qubitwise characteristic matrix in Eq. (\ref{qubitwise characteristic matrix}). Specifically, dividing the diagonals of $\mathbb{U}^Q$ by 2 and replacing the off diagonals as $\mathbb{U}^Q_{ij} \to \lceil 1 - \mathbb{U}^Q_{ij}/2 \rceil$, the uncertainty matrix becomes equal to the qubitwise characteristic matrix. An example network made up of an EPR source and a three-partite GHZ source together with its corresponding correlation matrices are illustrated in Fig.\ \ref{Example_five-qubit_network}. We now show that how the network topology can be inferred from the qubitwise uncertainty matrix.
	
	\textit{Theorem 1.} For a quantum network $\mathcal{N}$ satisfying assumptions (A) and (C), the network's topology is uniquely determined by its qubitwise uncertainty matrix $\mathbb{U}^Q(\mathcal{N})$.
	\begin{proof}
		First, since qubits in each node are measured individually, it is natural that the indices of qubits sent to each node are known. That is, for every qubit, the node $C_j$ to which it is sent is unambiguously identified. Next, recall that the two-qubit uncertainty values for $|\phi\rangle\langle\phi|$, $\tau_2$ and $\mathbb{I}_4/4$ are 0, 1 and 2, respectively. Hence, for any qubit $Q_i$, if $\mathbb{U}^Q_{ij} = 0$, then $Q_i$ and $Q_j$ form an EPR source, while $Q_i$ along with all such $Q_k$ satisfying $\mathbb{U}^Q_{ik} = 1$ constitute a multipartite GHZ source. Finally, by iterating over all rows of $\mathbb{U}^Q(\mathcal{N})$, one obtains the set of sources and their associated qubit indices, thereby uniquely identifying all links $(A_i, C_j)$ of $\mathcal{N}$.
	\end{proof}
	
	Concerning the ubiquitous noises in reality, which inevitably lead to increased uncertainties, one can instead apply a coarse-grained description to group the qubits\ \cite{chen_inferring_2023}. This is detailed in the following protocol, which incorporates the procedure for evaluating qubitwise uncertainty matrix and quantifying entanglement between two qubits.

	\begin{breakablealgorithm}
		\renewcommand{\thealgorithm}{}
		\caption{Network topology inference with uncertainties}
			\begin{enumerate}[(1)]
				\item All nodes in the network $\mathcal{N}$ measure observable $X$ on their received qubits. After repeating the measurement sufficient times, the nodes utilize classical communication to determine the probability distributions $\mathbb{P}(x_i)$ and $\mathbb{P}(x_i, x_j)$ for all $i=1,\dots,N_Q$ and $j \neq i$, where $x_i$ denotes the measurement outcome of qubit $Q_i$.
				\item All nodes measure observable $Z$ on their received qubits. This can be achieved by applying a Hadamard transformation to each qubit before the $X$ measurement. The probability distributions $\mathbb{P}(z_i)$ and $\mathbb{P}(z_i, z_j)$ are obtained likewise.
				\item The one-qubit and two-qubit uncertainties are calculated via $H_u(Q_i) = H(\mathbb{P}(x_i)) + H(\mathbb{P}(z_i)) $ and $H_u(Q_i|Q_j) = H(\mathbb{P}(x_i,x_j)) - H(\mathbb{P}(x_j)) + H(\mathbb{P}(z_i,z_j)) - H(\mathbb{P}(z_j))$.
				\item In the absence of a shared reference frame, each node optimizes the choice of MUBs and repeats steps (1)-(3). The minimum uncertainties are retained to construct $\mathbb{U}^Q(\mathcal{N})$.
				\item For $i=1,\dots,N_Q$:
					\begin{enumerate}[(a)]
						\item For $j$ satisfying $\mathbb{U}^Q_{ij}(\mathcal{N}) < 1$, the set $\lbrace i, j \rbrace$ is identified as an EPR source. Identical sets of qubit indices are counted only once. The distillable entanglement between $Q_i$ and $Q_j$ is lower bounded by $1-\mathbb{U}^Q_{ij}(\mathcal{N})$. 
						\item Let $K^i = \lbrace k^i | 1\leq \mathbb{U}^Q_{i k^i}(\mathcal{N}) < 2 \rbrace$, the set $\lbrace i \rbrace \cup K^i$ is identified as a multipartite GHZ source. Identical sets of qubit indices are counted only once.
					\end{enumerate}
				\item For each qubit $Q_i\ (i=1,\dots,N_Q)$, find the source set $A_m$ containing $i$, locate the node $C_n$ to which $Q_i$ is sent, and finally add the tuple $(A_m, C_n)$ to the set of links.
			\end{enumerate}
	\end{breakablealgorithm}
	Note that step (5) can be simplified by grouping the qubits that satisfy $\mathbb{U}^Q_{ij}(\mathcal{N}) < 2$ into one source. This modification categorizes all sources into a single class, yet ensures the protocol's applicability in scenarios with extreme interference.
	
	The central advantage of the proposed approach is that, it provides a quantitative estimation of entanglement in the quantum network. In real-life implementations, entanglement sources are susceptible to noise induced by imperfections or environments, quantification of entanglement is thus of particular relevance to quantum network quality assessments, especially in the absence of knowledge about the distributed states. According to Eq.\ (\ref{lower bound of distillable entanglement}), the off-diagonal term $\mathbb{U}^Q_{ij}(\mathcal{N})$ establishes a lower bound of the distillable entanglement $E_D (Q_i, Q_j)$ of qubit $Q_i$ and $Q_j$. Aside from bipartite entanglement, detection of multipartite entanglement using three-qubit uncertainty is feasible as well. The tripartite entanglement of formation $E_{3F}$ \cite{szalay_multipartite_2015} is lower bounded by Ref.\ \cite{schneeloch_quantifying_2020}
	\begin{equation}
		\begin{aligned}
			&E_{3F}(Q_i,Q_j,Q_k)  \\
			\ge &-H(Q_i | Q_j Q_k) - H(Q_j | Q_i Q_k) - H(Q_k | Q_i Q_j) - 2 \\
			\ge& 1 - H(X_i | X_j X_k) - H(Z_i | Z_j Z_k) - H(X_j | X_i X_k)\\
			&- H(Z_j | Z_i Z_k) - H(X_k | X_i X_j) - H(Z_k | Z_i Z_j),
			\label{tripartite entanglement detection}
		\end{aligned}
	\end{equation}
	where $H(Q_i | Q_j Q_k)$ is the conditional von Neumann entropy of system $Q_i$ given $Q_j$ and $Q_k$ together, and the second inequality is attained by invoking Eq.\ (\ref{EUR in presence of quantum memory}) and the data-processing inequality. Equation\ (\ref{tripartite entanglement detection}) tightly quantifies the tripartite entanglement of formation for three-partite GHZ state $(|000\rangle + |111\rangle)/\sqrt{2}$. The strength of this method in terms of white-noise tolerance is given in Ref.\ \cite{schneeloch_quantifying_2020}. For GHZ state mixed with white noise $p |\text{GHZ}_3\rangle\langle\text{GHZ}_3| + (1-p)\mathbb{I}_8 /8$, the conditional Shannon entropies in the last two lines of Eq.\ (\ref{tripartite entanglement detection}) are greater than 0 so long as $p>0.9406$. Notably, these conditional Shannon entropies can also be calculated by exploiting the same measurement strategies of two MUBs.
	
	What is more, the defined qubitwise uncertainty matrix has the potential for applications in network communication. For instance, the lower bound on the pairwise quantum key distribution capacitance of network $\mathcal{N}$ can be obtained from $\mathbb{U}^Q(\mathcal{N})$ according to Eq.\ (\ref{pairwise secret key}),
	\begin{equation}
		K_{\text{tot}} \ge \sum_{i<j} \max \{ 1 - \min\{ \mathbb{U}^Q_{ij}(\mathcal{N}), \mathbb{U}^Q_{ji}(\mathcal{N})\}, 0 \},
		\label{total key capcitance}
	\end{equation}
	where the minimum is taken considering the asymmetry of $\mathbb{U}^Q(\mathcal{N})$ in realistic experiments. Note that Eq.\ (\ref{total key capcitance}) actually quantifies the total key rates that can be extracted from the overall measurement data. Hence, the qubitwise matrix defined via uncertainty is more informative than its counterpart defined via mutual information or covariance.

	\subsubsection{Comparison and simulation of different approaches}\label{Comparison and simulation}
	Despite the operational meanings possessed by uncertainty, it demands two measurement settings in exchange for more extracted information, while the estimation of mutual information or covariance only needs one. In this sense, the measurement complexity and time complexity required to compute the uncertainty are doubled when compared to those of mutual information. Yet for covariance, it costs less time and fewer computational resources than uncertainty and mutual information in general, since the calculation of entropies involve logarithms. Moreover, noisy channels normally have a smaller impact on the value of covariance in contrast with entropic quantities, as demonstrated in Ref. \cite{chen_inferring_2023}.
	
	One may regard covariance as a preferable instrument to employ in network topology inference. We emphasize that covariance obtained in single basis (Eq.\ (\ref{covariance})) is not sufficient to detect entanglement, nor is mutual information (Eq.\ (\ref{mutual information})). Besides, though the covariance matrix obtained from local measurements allows semidefinite test of a given network topology \cite{aberg_semidefinite_2020}, it is not known what immediate information about the underlying network (such as the number or type of sources) the covariance matrix can provide. This renders the covariance-based approach hard to implement when plenty of alternative configurations need to be certified or large-scale quantum networks are under consideration. Nonetheless, not only qubitwise uncertainty in Eq.\ (\ref{two-qubit uncertainty}) enables bipartite entanglement quantification, but also the number of EPR sources can be counted by combining uncertainty and mutual information between nodes (see Sec.\ \ref{collective measurements} below). All three measures of correlations thus have their own advantages in different contexts.
	
	\begin{figure}[t]
		\centering
		\includegraphics[width=0.48\textwidth]{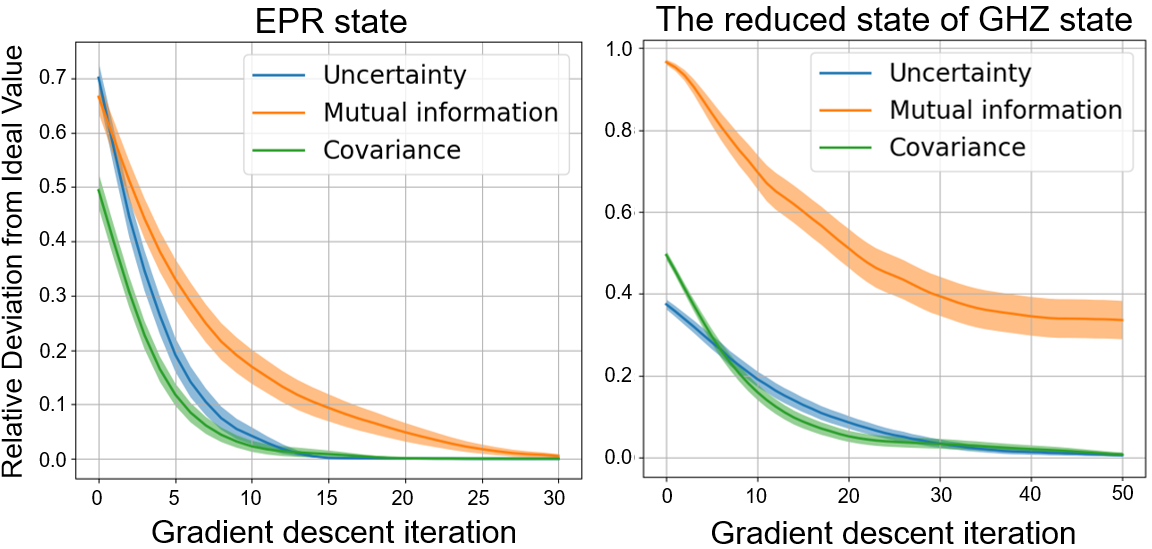}
		\caption{\label{Two-qubit_optimization}Variational quantum optimization of uncertainty, mutual information and covariance for EPR state (left panel) and the reduced state of GHZ state (right panel). Each plot shows the relative deviations of the correlations from their expected values averaged over 100 independent trials vs optimization step, with the shaded region corresponding to one standard error. The relative deviations of uncertainty and covariance are calculated by dividing the differences between their values to ideal values by half, since uncertainty ranges from 0 to 2, covariance ranges from -1 to 1, while the value scope of mutual information is 0 to 1. We take 10000 measurement shots to collect statistics. For EPR state, the step sizes for uncertainty, mutual information and covariance are 0.1, 0.1, and 0.2, whereas for the reduced state of GHZ state are 0.1, 0.25, and 0.3, respectively.}
	\end{figure}

	Before the comparison of experimental resource overheads, let us firstly introduce the variational quantum optimization method \cite{cerezo_variational_2021} used in numerical simulations. Recall that owing to the absence of shared reference frame across network nodes, mutual information and covariance require maximization while von Neumann entropies and uncertainties should be minimized. For uncertainty, the projectors that form the mutually unbiased qubit observables $X_i$ and $Z_i$ are expressed as
	\begin{align}
		| \mathbb{X}_{x_i}\rangle \langle \mathbb{X}_{x_i} | &= U(\vec{\theta}_i)^{\dagger} | x_i \rangle \langle x_i | U(\vec{\theta}_i), \label{X measurement}\\
		| \mathbb{Z}_{z_i}\rangle \langle \mathbb{Z}_{z_i} | &= U(\vec{\theta}_i)^{\dagger} H_i | z_i \rangle \langle z_i | H_i U(\vec{\theta}_i),
		\label{Z measurement}
	\end{align}
	where $x_i,z_i \in \{0, 1\}$, the unitary operator $U(\vec{\theta}_i)$ parameterized by $\vec{\theta}_i \in \mathbb{R}^3$ specifies the measurement direction of qubit $Q_i$ and $H_i$ is the Hadamard transformation on $Q_i$. To calculate uncertainties, the two observables of each qubit are chosen with uniform probability. The other quantities involve merely one observable. The gradient-descent algorithm \cite{chen_inferring_2023} is then applied to obtain the minimum of the cost functions
	\begin{align}
		H_{u}(Q_i | Q_j) =& \min_{\vec{\theta}_i, \vec{\theta}_j}\ H\left( \mathbb{P}(x_i, x_j | \vec{\theta}_i, \vec{\theta}_j)\right)  - H\left( \mathbb{P}(x_j | \vec{\theta}_j)\right) \notag \\
		& + H\left( \mathbb{P}(z_i, z_j | \vec{\theta}_i, \vec{\theta}_j)\right)  - H\left( \mathbb{P}(z_j | \vec{\theta}_j)\right) ,\\
		H_{u}(Q_i) =& \min_{\vec{\theta}_i}\ H\left( \mathbb{P}(x_i| \vec{\theta}_i)\right)   + H\left( \mathbb{P}(z_i | \vec{\theta}_i)\right)  .
	\end{align} 
	Parameters $\vec{\theta}_i$, $\vec{\theta}_j$ are updated in the direction opposite to gradients of the cost functions, towards approaching the optimal settings. The variational optimization framework is particularly effective in multiuser network scenarios as it resolves the problem of reference-frame alignment or suboptimal measurement choices. The following simulations are performed using the python software \texttt{qNetVO}: the Quantum Network Variational Optimizer \cite{qNetVO} under the \texttt{PennyLane} framework \cite{bergholm_pennylane_2022}.
	
	\begin{figure}[b]
		\centering
		\includegraphics[scale=0.32]{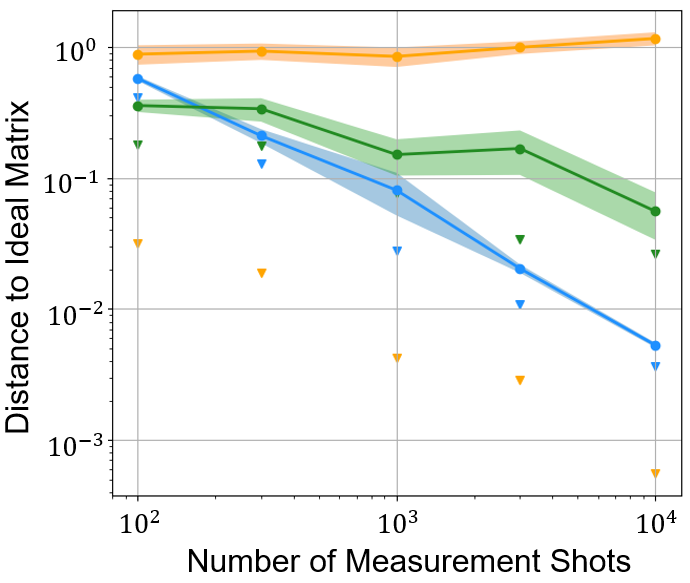}
		\caption{\label{Averaged distances}Averaged distances (circles with shaded standard error) and minimum distances (dashed triangles) of the optimized matrices across all trials as a function of the number of shots. We consider the five-qubit network in Fig.\ \ref{Example_five-qubit_network}. The performances of qubitwise uncertainty matrices (blue) and covariance matrices (green) fit the intuition that more shots lead to higher accuracy. Yet for qubitwise characteristic matrices (orange), the effect of shot noise is dominated by the probabilistic failure of mutual information optimization. The step sizes for uncertainty and covariance optimization are 0.1 and 0.3, respectively, while in qubitwise characteristic matrices, the von Neumann entropy entries are optimized with a step size of 0.1 and the mutual information entries are optimized with a step size of 0.25.}
	\end{figure}

	\begin{figure*}[t]
		\begin{center}
			\subfigure{\includegraphics[width=0.75\linewidth]{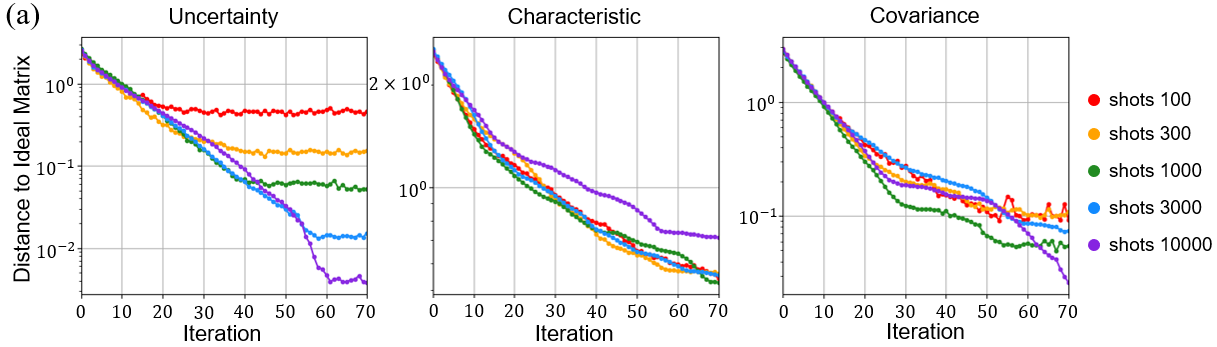}}
			\label{Distance_of_averaged_matrices}
			\vspace{-2mm}
			\subfigure{\includegraphics[width=0.75\linewidth]{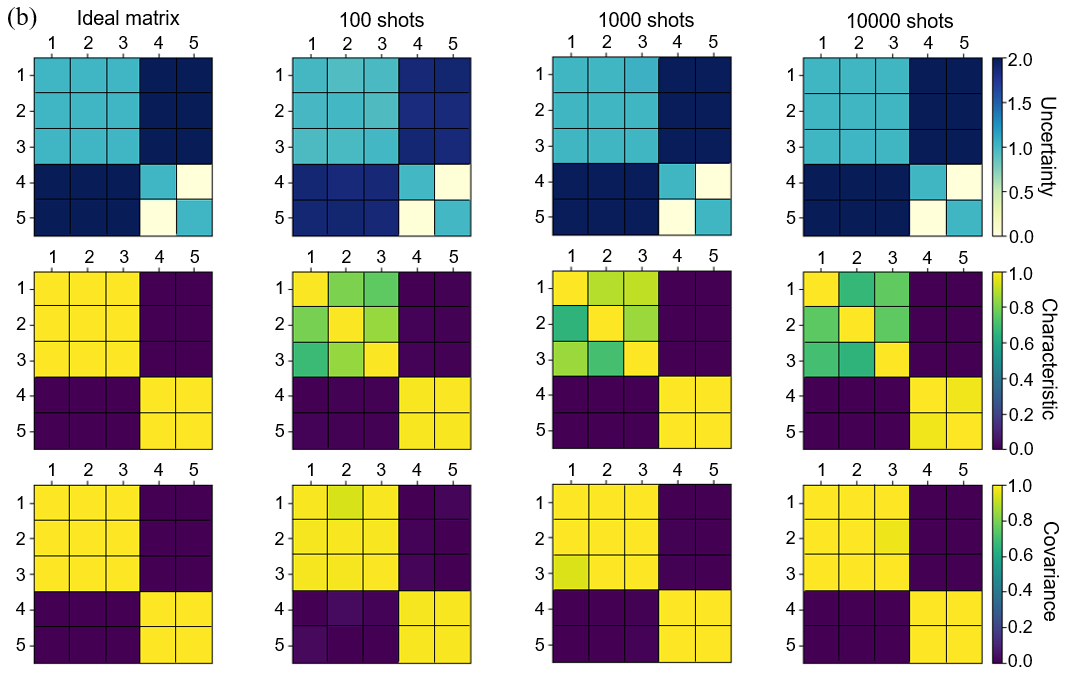}}
			\label{Matrix_heatmap}
			\caption{Variational optimization of correlation matrices for the network consisting of a three-qubit GHZ state and an EPR state. (a) Euclidean distances between the averaged matrices over 20 trials and the ideal matrices as a function of optimization steps. In each plot we show the optimization results under 100 shots (red), 300 shots (orange), 1000 shots (green), 3000 shots (blue), and 10 000 shots (purple). (b) Graphical depiction of optimized qubitwise uncertainty matrices (the first row), qubitwise characteristic matrices (the second row) and the covariance matrices (the third row) averaged over 20 trials. From left to right, each column plots the ideal matrices, the 100-shot matrices, the 1000-shot matrices and the 10000-shot matrices. Note that the diagonal terms of qubitwise uncertainty matrices are divided by 2. The figures are plotted using the same data as in Fig.\ \ref{Averaged distances}.}
		\end{center}
	\end{figure*}

	We now numerically investigate different approaches with respect to resource overheads, as reflected in optimization speed and sampling complexity. In each step of variational optimization, measurement statistics are collected via many shots. Thus, a smaller number of necessary optimization steps to achieve sufficiently strong correlations indicates a lower requirement for quantum resources, making the corresponding quantities more favorable in practical experiments. The optimization results of uncertainty, mutual information and covariance of $| \phi \rangle\langle \phi |$ and $\tau_2$ with random initial settings are shown in Fig. \ref{Two-qubit_optimization}. For both states, the optimization speed of uncertainty is faster than that of mutual information, and is comparable to that of covariance. Moreover, the nonconvergent mutual information of $\tau_2$ stems from probabilistic sticking in local minima during optimization, consistent with the results observed in Refs.\ \cite{chen_inferring_2023,qNetTI}. This highlights that the mutual information optimization for multipartite GHZ state exhibits a considerable failure probability and lower numerical stability in contrast with uncertainty and covariance.
	
	To show the sampling complexity of the network topology inference approaches, we compare their performances by varying the number of measurement shots to accumulate statistics. The performance of each approach is quantified by the Euclidean distance $d(M,M^\star) = \sqrt{\operatorname{tr}[(M - M^\star)^T (M - M^\star)]}$ between the optimized correlation matrix $M^\star$ and the ideal matrix $M$. The simulations are conducted in a noiseless five-qubit network where the first three qubits form a GHZ state and the remaining two qubits form an EPR state, with the ideal matrices displayed in Fig.\ \ref{Example_five-qubit_network}. 

	We optimize the qubitwise uncertainty matrix and the qubitwise characteristic matrix entrywise, while off diagonals of the covariance matrix are optimized according to Eq.\ (\ref{covariance}), and its diagonals (variances) are defined similarly. Averaged and minimum distances to ideal matrices across all 20 optimization trials versus the number of measurement shots are illustrated in Fig.\ \ref{Averaged distances}, taking 100, 300, 1000, 3000 and 10000 shots. For a fair comparison, the random initial settings of all correlation matrices are identical in each trial. Here we observe similar behavior for the qubitwise uncertainty matrix and covariance matrix, whose distances to their respective ideal matrices generally decrease when more shots are used. In contrast, for qubitwise characteristic matrix, the effect of probabilistic optimization failure significantly outweighs that of statistical shot noise. This results in the averaged distances largely independent of the number of shots, though the minimum distances properly descends. It is worth noting that uncertainty and covariance suffer from the local minima problem as well, but their empirical probabilities to get a failed optimization is greatly lower compared to mutual information by more than an order of magnitude.
	
	In Fig.\ \hyperref[Distance_of_averaged_matrices]{6(a)}, we plot the distances between the optimized matrices averaged over 20 trials to the ideal matrices at each optimization step. We find that the uncertainty-based approach exhibits better convergence and separation than covariance-based one, and it enables a faster optimization in low-shot regime. Distances of the averaged qubitwise characteristic matrices that are irrelevant of the number of shots are also shown here. Figure\ \hyperref[Matrix_heatmap]{6(b)} plots the values for all entries in the averaged matrices when 100, 1000, and 10 000 shots are used. The diagonals of qubitwise uncertainty matrices, i.e., the one-qubit uncertainty with twice the theoretical value of von Neumann entropy, are intentionally halved for a better visualization. One can see that 100 shots already suffice for valid network topology inference when noise is absent, in spite of its relatively large distance to the ideal matrix. The mutual information entries in three-qubit GHZ state are markedly spoiled due to local minima, thus one might incorrectly infer network topologies with high probability if inadequate trials were run. 
	
	From the numerical experiments on simulators, we see that the uncertainty-based approach requires fewer optimization trials and iterations for networks containing multipartite GHZ states. This makes it more reliable for inferring the networks despite its conceptual similarity between the compared quantities. In Appendix\ \hyperref[Inferring the network of a W state and a generalized EPR state]{A}, we further compare different approaches in a five-qubit network that includes two types of entangled states beyond GHZ: a W state and a generalized EPR state.

	\subsection{Probing networks from collective measurements}\label{collective measurements}
	It remains to consider the circumstance where qubits in each local node are measured in a collective manner, instead of being measured individually. This corresponds to the scenario where the user at each node does not perform individual discrimination on the received qubits. Here, we restrict the discussion to a noiseless scenario and to the networks satisfying assumptions (A)-(C). Motivated by the characteristic matrix in Eq.\ (\ref{characteristic matrix}), we aim to probe the network topology using the \textit{node-wise uncertainty matrix}
	\begin{align}
		&\mathbb{U}(\mathcal{N}) \notag \\
		=&\begin{pmatrix}
			H_{u}(C_1) & H_{u}(C_1 | C_2) & \cdots & H_{u}(C_1 | C_{N_C})\\
			H_{u}(C_2 | C_1) & H_{u}(C_2) & \cdots & H_{u}(C_2 | C_{N_C})\\
			\vdots & \vdots & \ddots & \vdots\\
			H_{u}(C_{N_C} | C_1) & H_{u}(C_{N_C} | C_2) & \cdots & H_{u}(C_{N_C})
		\end{pmatrix}.
		\label{node-wise uncertainty matrix}
	\end{align}
	Elements of the above matrix are defined as
	\begin{align}
		H_{u}(C_i) = &H(X_{i_1} X_{i_2} \dots X_{i_n}) + H(Z_{i_1} Z_{i_2} \cdots Z_{i_n}), \label{one-node uncertainty}\\
		H_{u}(C_i | C_j) = &H(X_{i_1} X_{i_2} \dots X_{i_n} | X_{j_1} X_{j_2} \dots X_{j_m} )\notag\\
		+& H(Z_{i_1} Z_{i_2} \dots Z_{i_n} | Z_{j_1} Z_{j_2} \dots Z_{j_m} ), \label{two-node uncertainty}
	\end{align}
	where $\{i_1, i_2, \dots, i_n\}$ is the set of indices of the qubits sent to node $C_i$, $H(X_{i_1} X_{i_2} \dots X_{i_n})$ is the Shannon entropy of outcomes of collective measurement $X_{i_1} X_{i_2} \dots X_{i_n}$ on all these qubits and the conditional Shannon entropy is $H(X_{i_1} X_{i_2} \dots X_{i_n} | X_{j_1} X_{j_2} \dots X_{j_m} ) = H(X_{i_1} \dots X_{i_n} X_{j_1}  \dots X_{j_m}) - H(X_{j_1} X_{j_2} \dots X_{j_m} )$. 
	
	Note that $X_{i_n}$ and $Z_{i_n}$ are still two mutually unbiased observables of qubit $Q_{i_n}$, so the measurement strategy is the same as when measurements at qubit level are available. Another resemblance is that all measurement directions normally need to be optimized in order to minimize the uncertainties as before. Remarkably, the number of optimizations for $\mathbb{U}(\mathcal{N})$ scales quadratically in the number of nodes. In comparison, the number of optimizations for the qubitwise uncertainty matrix $\mathbb{U}^Q(\mathcal{N})$ is typically much larger, as each node may receive multiple qubits.
	
	Analogous to the characteristic matrix, the diagonal term of node-wise uncertainty matrix also counts the number of qubits a node receives. Since the state of qubits in each node is a tensor product of $\mathbb{I}_2/2$, the uncertainty of their collective measurement outcomes will be
	\begin{align}
		&H(X_{i_1} X_{i_2} \cdots X_{i_n}) + H(Z_{i_1} Z_{i_2} \cdots Z_{i_n}) \notag \\
		=& H(X_{i_1}) + H(Z_{i_1}) + H(X_{i_2}) + H(Z_{i_2}) + \cdots + H(X_{i_n}) + H(Z_{i_n}) \notag \\
		=& H_{u}(Q_{i_1}) + H_{u}(Q_{i_2}) + \cdots + H_{u}(Q_{i_n})
	\end{align}
	due to the additivity of von Neumann entropy for independent subsystems. As the value of one-qubit uncertainty is just 2, one can see that $H_{u}(C_i) = 2n$ with $n$ being the number of qubits sent to node $C_i$, establishing a pragmatic interpretation of the uncertainty observed in a local node.
	
	Nevertheless, two distinct quantum network topologies may be indistinguishable solely from their node-wise uncertainty matrices. A primary example is the two networks in Fig.\ \ref{Example_networks}, wherein the joint state between any two nodes in $\mathcal{N}_a$ is $\rho_a = ( \mathbb{I}_2 \otimes |\phi\rangle \langle\phi| \otimes \mathbb{I}_2 ) / 4$, while the state for $\mathcal{N}_b$ is $\rho_b = (|00\rangle \langle00| + |11\rangle \langle11|)^{\otimes2} / 4$ \cite{chen_inferring_2023}. One can easily calculate that the minimum uncertainties between two nodes will be 2 for both networks, leading to identical node-wise uncertainty matrices
	\begin{equation}
		\mathbb{U}(\mathcal{N}_a) = \mathbb{U}(\mathcal{N}_b) = \begin{pmatrix}
			4 & 2 & 2\\
			2 & 4 & 2\\
			2 & 2 & 4
		\end{pmatrix}.
	\end{equation}
	In particular, uncertainties in $Z$ basis and $X$ basis are both 1 for $\rho_a$ on account of its symmetry. Yet for $\rho_b$, its uncertainty in $Z$ basis is 0 while the uncertainty in $X$ basis has a value of 2. Here one can notice that, it is the redundant information given by two bases that makes the network topologies unable to be uniquely determined.
	
	Most intriguingly, the uncertainties (Eqs. (\ref{one-node uncertainty}) and (\ref{two-node uncertainty})) can immediately provide additional information about the number of EPR sources shared between two nodes, if their measured mutual information (see Eq.\ (\ref{mutual information})) is evaluated as well. Subsequently, this enables a direct count of the total number of EPR sources in the quantum network by iterating through all pairs of nodes. Since EPR states are fundamental resources upon which are based plentiful quantum information processing tasks, the following theorem presents a useful tool for characterizing quantum networks.
	
	\textit{Theorem 2.}\label{Theorem 2} Let $C_i$ and $C_j$ be two arbitrary nodes in a noiseless network $\mathcal{N}$ satisfying assumptions (A)-(C). Given their uncertainties $H_{u}(C_i),\ H_{u}(C_j),\ H_{u}(C_i | C_j),\ H_{u}(C_j | C_i)$ and measured mutual information $I(C_i, C_j)$, the number of EPR sources shared between $C_i$ and $C_j$ is given by $N_{A, ij}^{\text{EPR}} = H_{u}(C_i) - H_{u}(C_i | C_j) - I(C_i, C_j) = H_{u}(C_j) - H_{u}(C_j | C_i) - I(C_i, C_j)$. The total number of EPR sources in network $\mathcal{N}$ is $N_{A}^{\text{EPR}} = \sum_{i \neq j} N_{A, ij}^{\text{EPR}}/2$.
	
	\begin{proof}
		Denote the set of nodes as $\mathbf{C} = \lbrace C_k | k = 1, 2, \dots, N_C \rbrace$ with $N_C$ being the total number of nodes in network $\mathcal{N}$. Let the set of nodes excluding $C_i$ and $C_j$ be $\mathbf{C}_{R} = \mathbf{C} \setminus \lbrace C_i,\ C_j \rbrace$, then all nodes are divided into three parts $\mathbf{C} = \lbrace C_i \rbrace \cup \lbrace C_j \rbrace \cup \mathbf{C}_{R}$. Because we are considering the correlations between $C_i$ and $C_j$, the sources that connect nodes solely in $\mathbf{C}_{R}$ have no influence on the two nodes of interest and are no longer considered in the following. Any one of the remaining sources must send qubits to at least two parts of $\mathbf{C}$, resulting in four possibilities.
		
		\begin{enumerate}[(1)]
			\item If a source connects $C_i$ and $C_j$, then the source is necessarily an EPR source. Denote the qubit sent to $C_i$ and $C_j$ as $Q_{i'}$ and $Q_{j'}$ respectively, one has $H_{u}(Q_{i'}|Q_{j'}) = H_{u}(Q_{j'}|Q_{i'}) =0,\ H_{u}(Q_{i'}) = H_{u}(Q_{j'}) = 2 .$
			\item If a source connects $C_i$, $C_j$ and node(s) in $\mathbf{C}_{R}$, then the source must be a multipartite GHZ source with at least three qubits and the joint state of $Q_{i'}$ and $Q_{j'}$ is $\tau_2 = (|00\rangle \langle 00| + |11\rangle \langle 11|) / 2$. One has $H_{u}(Q_{i'}|Q_{j'}) = H_{u}(Q_{j'}|Q_{i'}) =1,\ H_{u}(Q_{i'}) = H_{u}(Q_{j'}) = 2 .$
			\item If a source connects $C_i$ and node(s) in $\mathbf{C}_{R}$, then node $C_j$ will not receive any qubit from the source. In such a case, the source generates no measurement outcomes in node $C_j$. The joint Shannon entropy of outcomes in $C_i$ and $C_j$ contributed by this source is simply the entropy of outcomes in $C_i$, for either $Z$ or $X$ basis. By the definition of conditional Shannon entropy, the uncertainty of outcomes in $C_j$ given those in $C_i$ is zero. Conversely, as the qubit sent to $C_i$ is maximally mixed, the uncertainty of outcomes in $C_i$ given those in $C_j$ is just 2.
			\item If a source connects $C_j$ and node(s) in $\mathbf{C}_{R}$, then node $C_i$ will not receive any qubit from the source. Similar arguments yield that the uncertainty of outcomes in $C_i$ given those in $C_j$ is exactly 0, while the uncertainty of outcomes in $C_j$ given those in $C_i$ is 2.
		\end{enumerate}
		Since each source is independent to each other, the conditional Shannon entropy contributed by separate source is additive \cite{nielsen_quantum}. Denote the number of the above four types of sources as $N_{A,ij}^{\text{EPR}},\ N_{A,ij}^{\text{GHZ}},\ N_{A,ij}^{iR}$ and $N_{A,ij}^{jR}$, then the relations between the uncertainties and the numbers of sources is given by
		\begin{equation}
			\begin{cases}
				H_{u}(C_i) = 2 (N_{A,ij}^{\text{EPR}} + N_{A,ij}^{\text{GHZ}} + N_{A,ij}^{iR})\\
				H_{u}(C_j) = 2 (N_{A,ij}^{\text{EPR}} + N_{A,ij}^{\text{GHZ}} + N_{A,ij}^{jR})\\
				H_{u}(C_i | C_j) = N_{A,ij}^{\text{GHZ}} + 2N_{A,ij}^{iR}\\
				H_{u}(C_j | C_i) = N_{A,ij}^{\text{GHZ}} + 2N_{A,ij}^{jR}.
			\end{cases}
			\label{uncertainties and sources}
		\end{equation}
		Unfortunately, there is no unique solution to Eq.\ (\ref{uncertainties and sources}). If $(N_{A,ij}^{\text{EPR}},\ N_{A,ij}^{\text{GHZ}},\ N_{A,ij}^{iR},\ N_{A,ij}^{jR})$ is a solution to the observed uncertainties, so is $(N_{A,ij}^{\text{EPR}}+2,\ N_{A,ij}^{\text{GHZ}}-1,\ N_{A,ij}^{iR}-1,\ N_{A,ij}^{jR}-1)$. Therefore, two distinct network topologies can produce the same node-wise uncertainty matrix.
		
		Moreover, following Lemma 3 in Ref.\ \cite{chen_inferring_2023}, the measured mutual information between two nodes counts the number of sources they share, including both EPR sources and GHZ sources. That is, one has an additional equation $I(C_i, C_j) = N_{A,ij}^{\text{EPR}} + N_{A,ij}^{\text{GHZ}}$. As a result, given $H_{u}(C_i),\ H_{u}(C_j),\ H_{u}(C_i | C_j),\ H_{u}(C_j | C_i)$ and $I(C_i, C_j)$, the tuple $(N_{A,ij}^{\text{EPR}},\ N_{A,ij}^{\text{GHZ}},\ N_{A,ij}^{iR},\ N_{A,ij}^{jR})$ that generates the observed five quantities is uniquely determined as
		\begin{equation}
			\begin{cases}
				N_{A,ij}^{iR} = H_{u}(C_i)/2 - I(C_i, C_j)\\
				N_{A,ij}^{jR} = H_{u}(C_j)/2 - I(C_i, C_j)\\
				N_{A,ij}^{\text{GHZ}} = 2I(C_i, C_j) + H_{u}(C_i | C_j) - H_{u}(C_i) \\
				\qquad \ \ = 2I(C_i, C_j) + H_{u}(C_j | C_i) - H_{u}(C_j).\\
				N_{A,ij}^{\text{EPR}} = H_{u}(C_i) - H_{u}(C_i | C_j) - I(C_i, C_j) \\
				\qquad \ \ = H_{u}(C_j) - H_{u}(C_j | C_i) - I(C_i, C_j)\\
			\end{cases}
		\label{Counting EPR}
		\end{equation}
		
		Summing $N_{A,ij}^{\text{EPR}}$ over all $(i, j)$ with $i \neq j$ counts each EPR source twice. As a result, the total number of EPR sources presented in $\mathcal{N}$ is $N_{A}^{\text{EPR}} = \sum_{i \neq j} N_{A, ij}^{\text{EPR}}/2$.
	\end{proof} 
	
	As mentioned earlier, inferring the topology from the correlations between nodes is extremely hard for large networks, despite the one-to-one correspondence between the characteristic matrix $M(\mathcal{N})$ and the network $\mathcal{N}$. We believe a substantial impediment for network topology inference in previous approach is that, one cannot directly learn the numbers of diverse sources from the characteristic matrix. The above Theorem partially resolves the problem by counting the number of EPR sources, regardless of the complexity of networks. 
	
	Furthermore, if each source in a network distributes no more than four qubits, the number of three-partite GHZ sources and four-partite GHZ sources can also be learned. Denote the total number of the two types of GHZ sources as $N_{A}^{\text{3 GHZ}}$ and $N_{A}^{\text{4 GHZ}}$. The number of qubits in a network $\mathcal{N}
	$ is $2N_{A}^{\text{EPR}} + 3N_{A}^{\text{3 GHZ}} + 4N_{A}^{\text{4 GHZ}}$ and is easily obtained by summing over the diagonals of $M(\mathcal{N})$ or $\mathbb{U}(\mathcal{N})$. On the other hand, summing over the off-diagonals of $M(\mathcal{N})$ counts each three-partite source $2\times C_3^2$ times and counts each four-partite source $2\times C_4^2$ times, where $C_m^n$ is the binomial coefficient. That is to say, $\sum_{i \neq j} I (C_i, C_j) = 2N_{A}^{\text{EPR}} + 6N_{A}^{\text{3 GHZ}} + 12N_{A}^{\text{4 GHZ}}$. As $N_{A}^{\text{EPR}}$ is known according to Theorem \hyperref[Theorem 2]{2}, one is able to determine $N_{A}^{\text{3 GHZ}}$ and $N_{A}^{\text{4 GHZ}}$ by calculating both characteristic matrix and node-wise uncertainty matrix. This result is capable of reducing the computational complexity of potential network-reconstruction algorithms. We further extend the applicability of Theorem \hyperref[Theorem 2]{2} to noisy circumstances in Appendix\ \hyperref[Mitigating errors in a network]{B}.
	
	\section{Mitigating the inference error}\label{QEM}
	In real-world quantum networks, qubits are transmitted through channels with ineluctable noise, which deteriorates qualities of the distributed entanglement and induces biases in results of correlation estimations. To reduce the effects of physical noise on network topology inference, we incorporate two QEM methods, namely probabilistic error cancellation (PEC) and virtual distillation (VD), to the calculation of statistical correlations. Both methods involve postprocessing on measured probabilities and require only operations local to each node.
	\subsection{Probabilistic error cancellation method}\label{PEC}
	The essential idea of PEC is that one can effectively invert a noisy channel by stochastically inserting operations of its inverse \cite{temme_error_2017}. As a demonstration, consider the phase-damping noise acting on single qubit state $\rho$, the paradigmatic channel describing such a noisy process is \cite{nielsen_quantum}
	\begin{equation}
		\begin{aligned}
			&\mathcal{E}(\rho) = K_0 \rho K_0^\dagger +  K_1 \rho K_1^\dagger\\
			\text{ with } &K_0 = \begin{pmatrix} 1 & 0 \\ 0 & \sqrt{1-\gamma} \end{pmatrix},\ K_1 = \begin{pmatrix} 0 & 0 \\ 0 & \sqrt{\gamma} \end{pmatrix},
		\end{aligned}\label{phase damping channel}
	\end{equation}
	where $0 \le \gamma \le 1$ is the noise parameter. The inverse of $\mathcal{E}$ can be decomposed as
	\begin{align}
		&\mathcal{E}^{-1}(\rho) \notag \\
		=& \frac{1}{\sqrt{1-\gamma}} \left( \frac{1+\sqrt{1-\gamma}}{2}\rho - \frac{1-\sqrt{1-\gamma}}{2} \sigma_z \rho \sigma_z  \right) .
	\end{align}
	
	Due to the presence of a negative coefficient, the inverse channel $\mathcal{E}^{-1}$ is unable to be realized straight away. Nevertheless, one can equivalently implement the inverse by appending the Pauli-$Z$ gate after the noisy channel with probability $(1-\sqrt{1-\gamma}) / 2$, then combine the output statistics accordingly to recover the error-free measurement probabilities. To be exact, the ideal probability for $\rho$ to render outcome $\alpha$ can be acquired as
	\begin{equation}
		\begin{aligned}
			\operatorname{Tr}( \rho \Pi_{\alpha}) = \frac{1}{\sqrt{1-\gamma}} &\left\lbrace \frac{1+\sqrt{1-\gamma}}{2} \operatorname{Tr} \left[ \mathcal{E}(\rho) \Pi_{\alpha}\right] \right. \\
			& - \left. \frac{1-\sqrt{1-\gamma}}{2} \operatorname{Tr} \left[ \mathcal{Z} \circ \mathcal{E}(\rho) \Pi_{\alpha}\right] \right\rbrace,
		\end{aligned}
	\end{equation}
	where $\mathcal{Z}(\rho) = \sigma_z \rho \sigma_z$ is the operation of $\sigma_z$ gate and $\Pi_{a}$ is the projector corresponding to outcome $\alpha \in \{0, 1\}$.
	
	To show the feasibility of PEC in network topology inference approaches, we simulate its performance on an phase-damped EPR source. For simplicity, we assume that each qubit in the EPR source undergoes an independent phase-damping channel with the same noise parameter $\gamma$, yielding the density matrix $\frac{1}{2}| 00\rangle \langle 00 | + \frac{1-\gamma}{2}| 00\rangle \langle 11 | + \frac{1-\gamma}{2}| 11 \rangle \langle 00 | + \frac{1}{2}| 11 \rangle \langle 11 |$. The theoretical value of uncertainty for the noisy state can be written analytically as 
	\begin{equation}
		H_{u}(Q_1 | Q_2) = -\frac{\gamma}{2} \log \frac{\gamma}{2} - \frac{2-\gamma}{2}\log \frac{2-\gamma}{2},
		\label{theoretical value of phase damping}
	\end{equation}
	while the maximal mutual information and covariance are both 1 when $\sigma_z$ is measured.
	In order to calculate the two-qubit correlations, we apply the Pauli-$Z$ gate on each qubit with probability $(1-\sqrt{1-\gamma}) / 2$ before its measurement, and then obtain the joint measurement probability
	\begin{align}
		P(\vec{\alpha}) = \frac{1}{1-\gamma} &\left\lbrace \frac{(1+\sqrt{1-\gamma})^2}{4} \operatorname{Tr}\left[ \mathcal{E}_1 \circ \mathcal{E}_2(| \phi \rangle \langle \phi|) \Pi_{\vec{\alpha}}\right] \right.\notag \\
		+ & \frac{(1-\sqrt{1-\gamma})^2}{4} \operatorname{Tr}\left[ \mathcal{Z}_1 \circ \mathcal{Z}_2 \circ \mathcal{E}_1 \circ \mathcal{E}_2(| \phi \rangle \langle \phi|) \Pi_{\vec{\alpha}}\right]\notag\\
		- &\frac{\gamma}{4} \operatorname{Tr}\left[ \mathcal{Z}_1 \circ \mathcal{E}_1 \circ \mathcal{E}_2(| \phi \rangle \langle \phi|) \Pi_{\vec{\alpha}}\right] \notag\\
		- &\left.\frac{\gamma}{4} \operatorname{Tr}\left[ \mathcal{Z}_2 \circ \mathcal{E}_1 \circ \mathcal{E}_2(| \phi \rangle \langle \phi|) \Pi_{\vec{\alpha}}\right] \right\rbrace,
		\label{error-mitigated joint probabilities}
	\end{align}
	where $\mathcal{Z}_{1(2)}$ is the operation of $\sigma_z$ gate acted on the first (second) qubit of EPR state $| \phi \rangle \langle \phi|$, $\mathcal{E}_{1(2)}$ is the phase-damping channel in Eq.\ (\ref{phase damping channel}) acting on the first (second) qubit and $\vec{\alpha} \in \{00, 01, 10, 11\}$. The error-mitigated joint probabilities are used to estimate two-qubit uncertainty, mutual information and covariance, which are optimized with the variational framework in Sec.\ \ref{Comparison and simulation}.
	
\begin{figure}[t]
		\centering
		\includegraphics[scale=0.30]{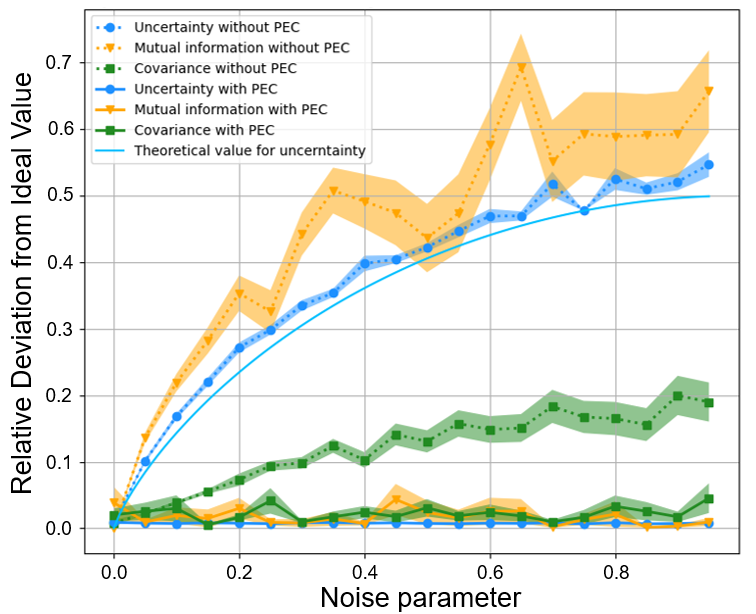}
		\caption{\label{PEC_phase_damping}The effect of phase-damping channels on evaluating the correlations of EPR state with and without PEC. We ran 50 independent trials with step size of 0.1 in the variational optimization of all curves. The number of optimization steps are 20 with PEC (solid lines) and 30 without PEC (dotted lines). We also plot the theoretical curve of uncertainty for the phase-damped EPR state, as in Eq. (\ref{theoretical value of phase damping}). The shaded regions indicate one standard error.}
	\end{figure}

	The numerical results with and without the application of PEC are illustrated in Fig.\ \ref{PEC_phase_damping}. We vary the noise parameter and run 50 independent trials employing \texttt{Pennylane}'s mixed-state simulator taking infinite shots. When PEC is not applied, the empirical values of uncertainty approach the theoretical noisy ones in Eq.\ (\ref{theoretical value of phase damping}), indicating its numerical stability. The drifts in mutual information and covariance without PEC result from the local minima problem as discussed in Sec.\ \ref{Comparison and simulation}, in that the noisy state approximates $\tau_2 = \frac{1}{2}\left( |00\rangle\langle00| + |11\rangle\langle11| \right)$ more when noise strength increases. Notably, we observe that PEC almost removes the deviations of all quantities, facilitating more accurate and reliable estimation of the correlation matrices. 
	
	As a remark, the elimination of biases in the PEC method comes at the price of complete characterization of the noise model, which is quite costly even for individual qubits. Noise modeling can be more intractable in network scenario, as multiple qubits transmitting through one physical link may experience correlated noise, yet PEC remains possible with the help of more sophisticated noise-characterization techniques \cite{guo_quantum_2022,strikis_learning-based_2021}. 
	
	Another point worth mentioning is that, since the error-mitigated probabilities in Eq.\ (\ref{error-mitigated joint probabilities}) are linearly combined, the effect of finite shot noise or unfaithful error modeling can induce fluctuations to terms in the combination. This can bring about negative probabilities during optimizations. The maximum-likelihood estimation technique \cite{james_measurement_2001} can be employed to address this negativity problem in entropy calculations. Alternatively, one can solve a constrained optimization problem to obtain the physical probability distribution that best approximates the mitigated unphysical one \cite{geller_rigorous_2020}, as detailed in Appendix\ \hyperref[Mitigating errors in a network]{B}.
	
	\subsection{Virtual distillation method}\label{Virtual distillation method}
	Different from PEC, VD is a representative of error-agnostic QEM techniques with no dependence on prior knowledge of the specific noise. The word ``virtual" reflects that the method does not actually prepare a purified state as in conventional distillation protocols \cite{bravyi_universal_2005}, instead it utilizes additional copies of the state to enhance the accuracy of its expectation values. Without loss of generality, consider a noisy quantum state $\rho = (1 - \epsilon) \rho_{\text{ideal}} + \epsilon \rho_{\text{error}}$, where $\rho_{\text{error}}$ is a state orthogonal to the ideal state $\rho_{\text{ideal}}$ and $0\le\epsilon\le1$ quantifies the magnitude of noise. By calculating the expectation values with respect to the $M$th degree state $\rho^M / \operatorname{Tr}(\rho^M)$, one can exponentially suppress the relative weights of the erroneous components so long as $\rho_{\text{ideal}}$ is the dominant component of $\rho$. Therefore, the objective of VD is to evaluate the error-mitigated expectation value of an observable of interest $O$, i.e.,
	\begin{equation}
		\langle O \rangle_{\text{VD}} = \frac{\operatorname{Tr}(O\rho^M)}{\operatorname{Tr}(\rho^M)}.
		\label{error-mitigated expectation value}
	\end{equation}
	
	One typical way to estimate Eq.\ (\ref{error-mitigated expectation value}) is to make use of the equality \cite{ekert_direct_2002}
	\begin{equation}
		\operatorname{Tr}(O \rho^M) = \operatorname{Tr}(O^{\mathbf{i}} S^{(M)} \rho^{\otimes M}) = \operatorname{Tr}(O^{(M)}S^{(M)} \rho^{\otimes M}),
		\label{Equality for VD}
	\end{equation}
	where $O^{\mathbf{i}}$ is the observable $O$ acting on the arbitrarily chosen $i$th copy of $\rho$, $O^{(M)}=\sum_{i=1}^M O^{\mathbf{i}}/M$ is the symmetrized observable and $S^{(M)}$ is the permutation operator on all copies acting as $S^{(M)} | \psi_1 \rangle | \psi_2 \rangle \cdots | \psi_M \rangle = | \psi_2 \rangle | \psi_3 \rangle \cdots | \psi_1 \rangle$. As such, one can obtain $\operatorname{Tr}(O\rho^M)$ and $\operatorname{Tr}(\rho^M)$ via measuring an ancilla qubit associated with controlled gates\ \cite{koczor_exponential_2021}. However, the implementation of controlled operations between even two copies can be quite demanding in near-term experiments and potentially introduce further errors.
	
	Meanwhile, the proposal of measuring multiple copies by diagonalization offers a strategy that is easy to use \cite{huggins_virtual_2021}. Based on the diagonalization method, we propose a protocol to calculate error-mitigated two-qubit correlations of the second degree state without the need of entangled ancilla. The proposed protocol requires only entangling gates between each pair of individual qubits from two copies and is thereby more favorable in network scenario.
	
	The observation is that, only three expectation values $\langle Z_i \rangle_{\text{VD}},\ \langle Z_j \rangle_{\text{VD}}$ and $\langle Z_i Z_j \rangle_{\text{VD}}$ are needed to calculate the error-mitigated joint probabilities of qubits $Q_i$ and $Q_j$,
	\begin{equation}
		\begin{aligned}
			P(mn)_{\text{VD}} = &\frac{1}{4} \left[1 + (-1)^m \langle Z_i \rangle_{\text{VD}} + (-1)^n \langle Z_j \rangle_{\text{VD}} \right. \\
			& \left. + (-1)^{m+n} \langle Z_i Z_j \rangle_{\text{VD}} \right],\quad \forall m,n \in \{0,1\}.  
		\end{aligned}
		\label{error-mitigated VD probabilities}
	\end{equation}
	Here, $Z_{i(j)}$ denotes the Pauli-$Z$ operator on qubit $Q_{i(j)}$, and $Z_i Z_j$ is the collective measurement operator. Note that measurements in other bases can be achieved by appending single-qubit rotations before the VD procedure, making the protocol compatible with the variational optimization framework in Eqs.\ (\ref{X measurement}) and (\ref{Z measurement}). In the following discussions, we focus on the reduced state $\rho_{ij}$ of $Q_i$ and $Q_j$.
	
	To start with, we show how to obtain the values of $\operatorname{Tr}(Z_i^{(2)} S^{(2)} \rho_{ij}^{\otimes 2})$, $\operatorname{Tr}(Z_j^{(2)} S^{(2)} \rho_{ij}^{\otimes 2})$ and $\operatorname{Tr}( S^{(2)} \rho_{ij}^{\otimes 2})$ in a simultaneous fashion \cite{huggins_virtual_2021}. First note that the permutation operator $S^{(2)}$ can be decomposed into local swap operations between the $i$th and $j$th qubits across two copies $S^{(2)}=S_i^{(2)} \otimes S_j^{(2)}$. Yet rather than swap gate, we instead apply the following unitary to each pair of qubits transversally before their measurement
	\begin{equation}
		B_k^{(2)} = \begin{pmatrix}
			1 & 0 & 0 & 0\\
			0 & \frac{1}{\sqrt{2}} & -\frac{1}{\sqrt{2}} & 0\\
			0 & \frac{1}{\sqrt{2}} & \frac{1}{\sqrt{2}} & 0\\
			0 & 0 & 0 & 1
		\end{pmatrix},\ k = i \text{ or } j.
		\label{diagonalization gate B}
	\end{equation}
	This unitary diagonalizes $S^{(2)}, Z_i^{(2)} S^{(2)}$ and $Z_j^{(2)} S^{(2)}$ at the same time. To be specific, we have the individual factors of the form
	\begin{align}
		&B_k^{(2)}S_k^{(2)}B_k^{(2)\dagger} \notag \\
		&= (|00\rangle\langle00| - |01\rangle\langle01| + |10\rangle\langle10| + |11\rangle\langle11|)_{kk^{\prime}}, \label{diagonalizing permutation}\\
		&B_k^{(2)} Z_{k}^{(2)} S_k^{(2)} B_k^{(2)\dagger} = (|00\rangle\langle00| - |11\rangle\langle11|)_{kk^{\prime}},
	\end{align}
	where $k^\prime$ refers to the $k$th qubit in the second copy.
	Consequently, one can linearly combine the probabilities measured in computational basis to obtain $\operatorname{Tr}(Z_i^{(2)} S^{(2)} \rho_{ij}^{\otimes 2})$ by noting that
	$\operatorname{Tr}(Z_i^{(2)} S^{(2)} \rho^{\otimes 2}) = \operatorname{Tr}(B^{(2)} Z_i^{(2)} S^{(2)} B^{(2)\dagger} B^{(2)} \rho^{\otimes 2} B^{(2)\dagger})$ where $B^{(2)} = B_i^{(2)} \otimes B_j^{(2)}$. In a similar manner, $\operatorname{Tr}(Z_j^{(2)} S^{(2)} \rho_{ij}^{\otimes 2})$ and $\operatorname{Tr}( S^{(2)} \rho_{ij}^{\otimes 2})$
	can be calculated using the same probability distribution. We thus attain $\langle Z_i \rangle_{\text{VD}}$ and $ \langle Z_j \rangle_{\text{VD}}$ according to Eq.\ (\ref{error-mitigated expectation value}).
	
	Our next goal is to capture $\operatorname{Tr}(Z_i^{\mathbf{1}} Z_j^{\mathbf{1}} S^{(2)} \rho_{ij}^{\otimes 2})$. Although the operator $Z_i^{\mathbf{1}} Z_j^{\mathbf{1}} S^{(2)}$ is non-Hermitian, it still admits a factorization into a tensor product of diagonal operators acting on separate pairs of qubits, i.e.,
	\begin{equation}
		\begin{aligned}
			&D^{(2)} Z_i^{\mathbf{1}} Z_j^{\mathbf{1}} S^{(2)} D^{(2)\dagger}\\
			=&(|00\rangle\langle00| +\mathrm{i} |01\rangle\langle01| - \mathrm{i} |10\rangle\langle10| - |11\rangle\langle11|)_{ii^{\prime}}\\
			\otimes & (|00\rangle\langle00| +\mathrm{i} |01\rangle\langle01| - \mathrm{i} |10\rangle\langle10| - |11\rangle\langle11|)_{jj^{\prime}},
			\label{expansion of ZiZj}
		\end{aligned}
	\end{equation}
	where
	\begin{align}
		&D^{(2)} = D_i^{(2)} \otimes D_j^{(2)},\\
		&D_k^{(2)} = \begin{pmatrix}
			1 & 0 & 0 & 0\\
			0 & \frac{1}{\sqrt{2}} & -\frac{\mathrm{i}}{\sqrt{2}} & 0\\
			0 & \frac{1}{\sqrt{2}} & \frac{\mathrm{i}}{\sqrt{2}} & 0\\
			0 & 0 & 0 & 1
		\end{pmatrix} = B_k^{(2)} \begin{pmatrix}
			1 & 0 & 0 & 0\\
			0 & 1 & 0 & 0\\
			0 & 0 & \mathrm{i} & 0\\
			0 & 0 & 0 & 1
		\end{pmatrix}.
		\label{diagonalization gate D}
	\end{align}
	Likewise, one combines the probabilities measured in computational basis after the action of the diagonalization gate $D_{k}^{(2)}$ on each pair of qubit, then divides the resulting $\operatorname{Tr}(Z_i^{\mathbf{1}} Z_j^{\mathbf{1}} S^{(2)} \rho_{ij}^{\otimes 2})$ by $\operatorname{Tr}( S^{(2)} \rho_{ij}^{\otimes 2})$ (obtained in the previous step) to get $\langle Z_i Z_j \rangle_{\text{VD}}$. The imaginary terms in the expansion of Eq.\ (\ref{expansion of ZiZj}) have no influence on the expectation value as long as the two copies are identical. Note that the qubits labeled by $i$ and $j$ can undergo different noise channels in principle.
	
	As a final step, the error-mitigated probabilities are evaluated by resorting to Eq.\ (\ref{error-mitigated VD probabilities}), which are further used to calculate the uncertainty in Eq.\ (\ref{two-qubit uncertainty}) or any other two-qubit correlations. By applying the diagonalization gates $B_k^{(2)}$ or $D_k^{(2)}$ to every pair of qubits across two copies of the network state, one is able to construct the correlation matrices that serve as the basis for network topology inference.
		
	\begin{figure}[!b]
		\centering
		\includegraphics[scale=0.37]{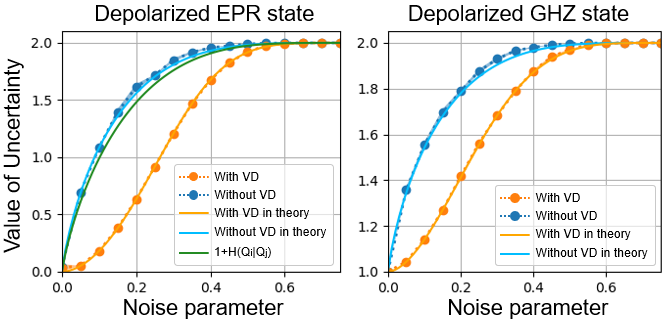}
		\caption{\label{VD_varying_noise}Optimized values of uncertainty in the unmitigated noisy states (blue) and the states processed by VD with two copies (orange) vs the noise parameter. We apply depolarizing noise to each qubit of EPR state (left panel) and the reduced state of GHZ state $\tau_2$ (right panel). Measurement statistics are collected via infinite shots, as in the case of PEC. We perform 20 trials of optimization with 20 steps for each curve and compare the simulation results with theory (solid lines). The lower bound of uncertainty $1+H(Q_i|Q_j)$ (green solid lines) is also plotted for EPR state, while it coincides with the achievable uncertainty for $\tau_2$. Noise-dependent step sizes are chosen to achieve optimal performance. The step sizes of EPR state, given depolarizing probability $\gamma$, are 0.05+$\gamma$ with VD and 0.05+$\gamma/2$ without VD, while those of $\tau_2$ are 0.2+2$\gamma$ with VD and 0.3+$\gamma$ without VD.}
	\end{figure}

	We numerically demonstrate the application of VD to reduce the error in uncertainty. We consider EPR state $|\phi\rangle\langle\phi|$ and the reduced state of GHZ state $\tau_2$, each qubit of which undergoes independent single-qubit depolarizing channels modeled by the following Kraus operators \cite{nielsen_quantum}:
	\begin{equation}
		\begin{aligned}
			& K_0 = \sqrt{1-\gamma}\ \mathbb{I}_2, K_1 = \sqrt{\frac{\gamma}{3}} \sigma_x,\\
			& K_2 = \sqrt{\frac{\gamma}{3}} \sigma_y, K_3 = \sqrt{\frac{\gamma}{3}} \sigma_z,
			\label{Depolarizing Kraus operators}
		\end{aligned}
	\end{equation}
	where $\gamma$ ranging from 0 to 3/4 is the depolarization probability. Then, the noisy states can be compactly written as 
	\begin{equation}
		\mathcal{E}(\rho) = \left( 1 - \frac{4\gamma}{3} \right)^2 \rho + \frac{6\gamma - 4 \gamma^2}{9} \mathbb{I}_4. 
	\end{equation}
	The measurement bases minimizing the uncertainties can be simply chosen as $\sigma_x^{\otimes2}$ and $\sigma_z^{\otimes2}$, one can analytically calculate the theoretical values by utilizing Eqs.\ (\ref{error-mitigated expectation value}) and (\ref{error-mitigated VD probabilities}).

	Figure\ \ref{VD_varying_noise} presents two plots that show the impacts of VD across the noise regime. The lower bound of uncertainty set by the EUR in Eq.\ (\ref{EUR in presence of quantum memory}) is also plotted. One can see that values of uncertainty for both $|\phi\rangle\langle\phi|$ and $\tau_2$ grows slower with respect to the noise parameter and the numerical results fit well with the achievable values in theory. Note that in the context of network topology inference, once we observe a two-qubit uncertainty less than unity, $H_u(Q_i | Q_j) < 1$, these two qubits necessarily originate from an EPR source even in noisy regime. Particularly, the threshold noise parameter for certifying EPR source is extended from 0.0876 to 0.2640 with the help of the VD method, indicating a threefold improvement in depolarizing noise tolerance enjoyed by the mitigated state. 
	
	We emphasize that, despite the more accurate uncertainties, the amount of actual entanglement (see Eq.\ (\ref{lower bound of distillable entanglement})) would be overestimated by the QEM methods considered here as they aim at restoration of measurement statistics rather than ideal noise-free states. Nonetheless, the above-introduced VD procedures still enable bipartite entanglement detection for unmitigated states via the R{\'e}nyi entropy inequality $S_2(\rho_{ij}) \ge S_2(\rho_{i})$ which holds if $\rho_{ij}$ is separable \cite{horodecki_quantum_2009}, as the second-order R{\'e}nyi entropies $S_2(\rho) = -\log \operatorname{Tr}(\rho^2) = -\log \operatorname{Tr}(S^{(2)}\rho^{\otimes 2})$ can be concurrently acquired by virtue of Eq.\ (\ref{diagonalizing permutation}).
	
	\begin{figure}[b]
		\centering
		\includegraphics[scale=0.32]{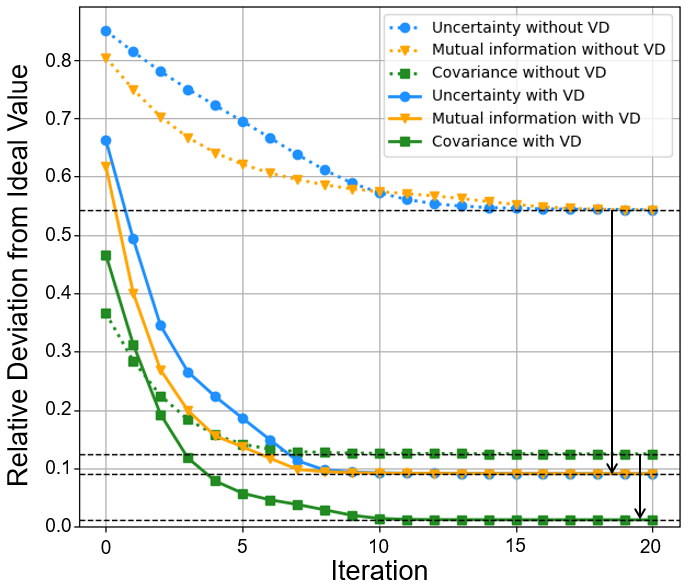}
		\caption{\label{VD_fixed_noise}Variational optimization of two-qubit correlations for depolarized EPR state with (solid) and without (dotted) the application of VD, fixing the noise parameter to 0.1. Each curve shows the relative deviations of the correlations from their ideal values vs the optimization iteration, averaged over 30 trials. Black arrows indicate the error reductions achieved by VD. We see that all curves converge to the corresponding theoretical values (dashed). The step sizes for uncertainty, mutual information, covariance are 0.15, 0.3, 0.3, respectively, when applying VD, and is 0.1, 0.3, 0.3 otherwise.}
	\end{figure}

	Additionally, we examine the number of iterations required to approach the theoretical limits for all three measures of correlation in Fig.\ \ref{VD_fixed_noise}, fixing the depolarization probability $\gamma$ of EPR state to be 0.1. We observe a faster convergence for the optimization of uncertainty and mutual information when VD method is employed. In contrast, the application of VD has a minor effect on the covariance optimization speed, and the error reduction is visibly smaller than that of the other quantities. As a proof-of-concept demonstration, these results suggest that the use of QEM can enhance the robustness and practicality of network topology inference by suppressing noise-induced errors of two-qubit correlations.

	Note that here we focus on the mitigation of errors in two-qubit correlations, while it is not straightforward to apply the diagonalization method to evaluate correlations between nodes (see Sec.\ \ref{collective measurements}). The challenge arises from the fact that factorization of the permuted multi-qubit observables is highly nontrivial \cite{huggins_virtual_2021}. To effectively reduce biases in outcomes of collective measurements, the recently proposed shadow distillation method \cite{seif_shadow_2023}, which replaces coherent operations on multiple copies by randomized single-qubit measurements, opens a feasible avenue for near-term application. In Appendix \hyperref[Mitigating errors in a network]{B}, we numerically demonstrate shadow distillation in both cases of qubit measurement and collective measurement.
	
	\section{Conclusion}\label{Conclusion}
	In this work, we have introduced an approach to infer the topology of quantum networks with entropic uncertainty by performing only two local qubit measurements. We have also incorporated the powerful quantum error-mitigation methodology into the framework to tackle the problem of noise. The proposed qubitwise matrices enable simple reconstruction of topologies and quantification of entanglement between arbitrary pair of qubits with no prior information about the network under consideration. In another physically relevant scenario where qubits in each local node are collectively measured, the uncertainty between two nodes can directly calculate the number of shared bipartite sources when used in conjunction with their mutual information. This result highlights the profound influence of entropic quantities on quantum network characterization. We simulate and compare approaches based on different measures of correlation in a five-qubit network and find that uncertainty-based approach exhibits higher numerical stability and faster convergence during variational optimization. 
	
	Furthermore, we have demonstrated the elimination of errors in statistical correlation estimations and a threefold depolarizing noise tolerance in certifying bipartite entanglement sources, benefited from probabilistic error cancellation and virtual distillation, respectively. The error-mitigating operations remain local to each node, providing a noise-resilient way of experimental topology inference for quantum network. This also furnishes a useful application of error-mitigation techniques beyond the usual realm of quantum computation. The approach is within reach using existing technology and is scalable to larger networks.
	
	Future work is expected to focus on relaxing the GHZ state-preparation assumption for applications in more intricate quantum networks. It would be highly desirable to generalize the analysis to sources distributing cluster states, which lie at the core of one-way quantum computation \cite{raussendorf_one-way_2001}, or other states that exhibit quantum advantages for realistic tasks. Exploring the extension to uncertainty between three qubits or three nodes can also be a fruitful direction. Moreover, exploiting \textit{a priori} information would offer fascinating possibilities to the uncertainty-based approach. Leveraging state-dependent uncertainty relations \cite{bergh_experimentally_2021} in the proposed approach allows tighter quantification of entanglement and, besides that, potentially leads to delicate tests tailored to network topologies.
	
	\section{Acknowledgments}
	We thank Xiongfeng Ma and Xiao-Hui Bao for enlightening discussions and valuable suggestions. This work has been supported by the National Natural Science Foundation of China (Grants No. 62375252, No. 62031024, No. 11874346, and No. 12174375), the National Key R\&D Program of China (Grant No. 2019YFA0308700), the Anhui Initiative in Quantum Information Technologies (Grant No. AHY060200), the Innovation Program for Quantum Science and Technology (Grant No. 2021ZD0301100, No. 2021ZD0301400), and the Fundamental Research Funds for the Central Universities.
	
	\section{Data availability}
	The data that support the findings of this article are openly available \cite{codeRepo}.
	
	\section*{Appendix A: Inferring the network of a W state and a generalized EPR state}\label{Inferring the network of a W state and a generalized EPR state}
	In this Appendix, we consider a five-qubit quantum network that contains two types of entangled states beyond GHZ---the $W$ state $|W\rangle_{123} = (|001\rangle + |010\rangle + |100\rangle)/\sqrt{3}$ and the generalized EPR state $|\text{EPR}^\prime\rangle_{45} = \cos(\pi/8)|00\rangle + \sin(\pi/8)|11\rangle$. 
	
	We first analytically analyze the correlation matrices based on different quantities. For the three-partite $W$ state, its two-qubit reduced state is $W_{12} = |00\rangle\langle 00|/3 + 2|\psi^+\rangle\langle \psi^+|/3$ with $|\psi^+\rangle = (|01\rangle+|10\rangle)/\sqrt{2}$. When both qubits are measured in Pauli $\sigma_x$ basis, the measured mutual information and covariance reach their maximum \cite{chen_inferring_2023}
	\begin{equation}
		I(Q_1, Q_2) \approx 0.3500 \equiv I_W,\ C(Q_1, Q_2) = 2/3.
	\end{equation}
	In order to attain the minimum entropic uncertainty $H_u(Q_1|Q_2) = 1.297 \equiv U_W$, the two MUBs for one qubit are \begin{align}
			&\left\lbrace \cos \frac{\pi}{8}|0\rangle + \sin \frac{\pi}{8} |1\rangle, -\sin \frac{\pi}{8}|0\rangle + \cos \frac{\pi}{8} |1\rangle \right\rbrace,\\
			&\left\lbrace \cos \frac{\pi}{8}|+\rangle - \sin \frac{\pi}{8} |-\rangle, \sin \frac{\pi}{8}|+\rangle + \cos \frac{\pi}{8} |-\rangle \right\rbrace,
		\end{align}
	while those for the other qubit are obtained by replacing the angles in the trigonometric functions with $3\pi/8$. Here $|+\rangle = (|0\rangle + |1\rangle)/\sqrt{2}$ and $|-\rangle = (|0\rangle - |1\rangle)/\sqrt{2}$. It is worth noting that when the MUBs of both qubits are $\sigma_z$ and $\sigma_x$, the entropic uncertainty of $W_{12}$ takes a suboptimal value of 1.317, which is slightly larger than the global minimum.

	\begin{figure*}[bt]
		\begin{center}
			\subfigure{\includegraphics[width=0.75\linewidth]{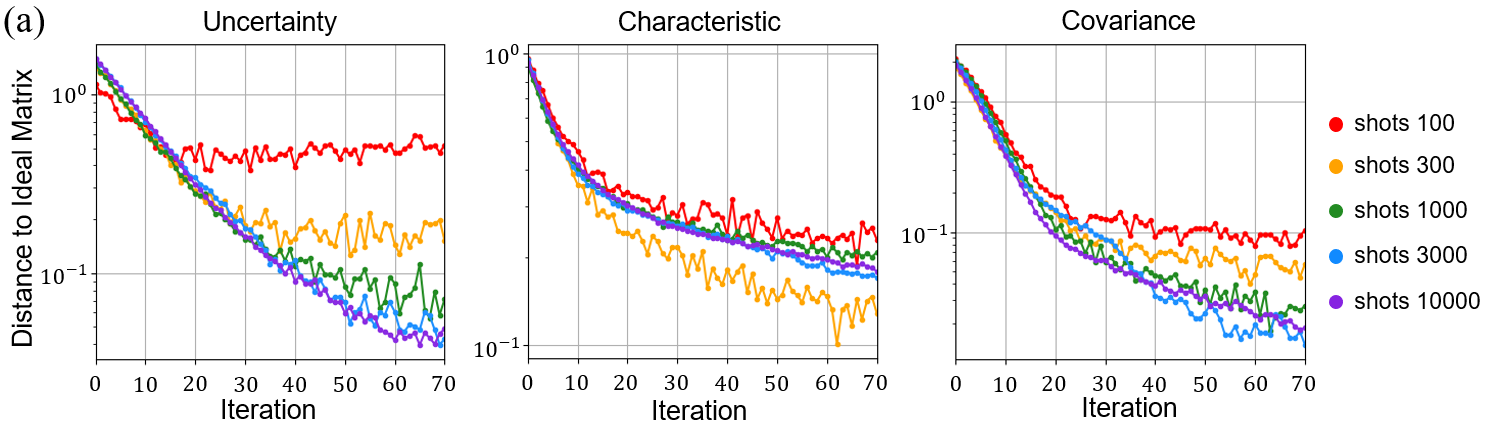}}
			\label{Distance_of_averaged_matrices_W}
			\vspace{-2mm}
			\subfigure{\includegraphics[width=0.75\linewidth]{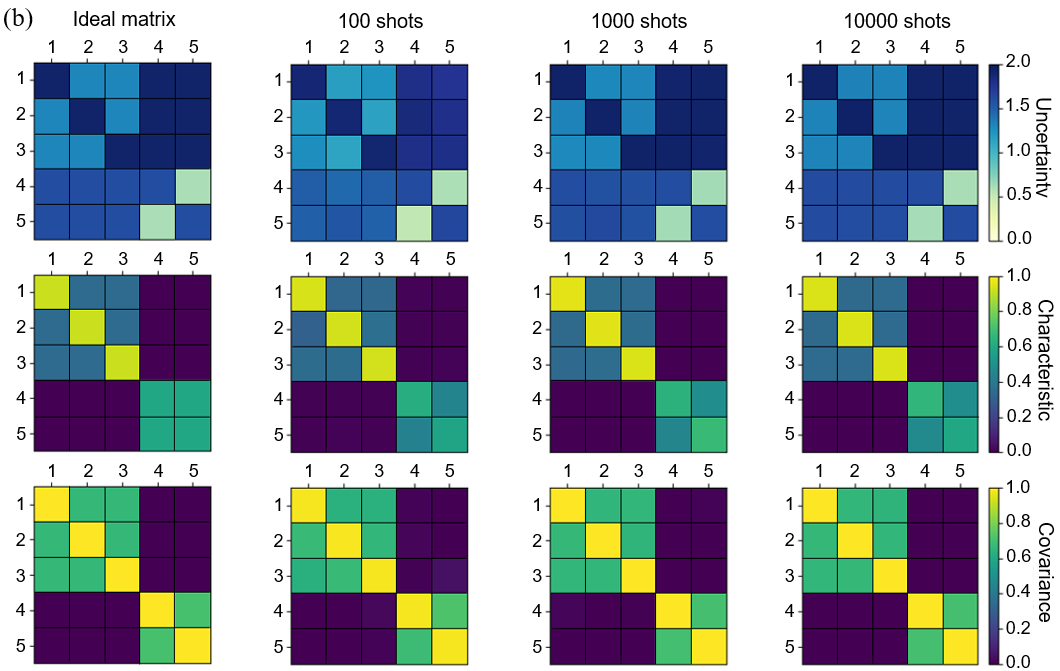}}
			\label{Matrix_heatmap_W}
			\caption{Variational optimization of correlation matrices for the network consisting of a three-qubit $W$ state and a generalized EPR state. (a) Euclidean distances between the averaged matrices over 20 trials and the ideal matrices as a function of optimization steps. In each plot we show the optimization results under 100 shots (red), 300 shots (orange), 1000 shots (green), 3000 shots (blue), and 10 000 shots (purple). (b) Graphical depiction of optimized qubitwise uncertainty matrices (the first row), qubitwise characteristic matrices (the second row) and the covariance matrices (the third row) averaged over 20 trials. From left to right, each column plots the ideal matrices, the 100-shot matrices, the 1000-shot matrices and the 10 000-shot matrices. The step sizes for uncertainty and covariance optimization are 0.1 and 0.2, respectively, while in qubitwise characteristic matrices, the von Neumann entropy entries are optimized with a step size of 0.2 and the mutual information entries are optimized with a step size of 0.3.}
		\end{center}
	\end{figure*}

	As for the generalized EPR state $|\text{EPR}^\prime\rangle_{45}$, the maximum mutual information between the two qubits is $2-[(2-\sqrt{2}) \log(2-\sqrt{2}) + (2+\sqrt{2}) \log(2+\sqrt{2})]/4 \approx 0.6009 \equiv E_G$ when both qubits are measured in $\sigma_z$ basis. The maximum covariance is $\sqrt{2}/2$ when both qubits are measured in $\sigma_x$ basis, while the minimum uncertainty equals $E_G$ with the MUBs being $\sigma_z$ and $\sigma_x$.
	
	By measuring the qubits corresponding to the remaining elements of the matrix in the $\sigma_z$ and $\sigma_x$ bases, we obtain the following ideal mutual information matrix
	\begin{equation}
		M^Q = \begin{pmatrix}
			E_W & I_W & I_W & 0 & 0\\
			I_W & E_W & I_W & 0 & 0\\
			I_W & I_W & E_W & 0 & 0\\
			0 & 0 & 0 & E_G & E_G\\
			0 & 0 & 0 & E_G & E_G
		\end{pmatrix},
	\end{equation}
	and ideal covariance matrix
	\begin{equation}
		C^Q = \begin{pmatrix}
			1 & 2/3 & 2/3 & 0 & 0\\
			2/3 & 1 & 2/3 & 0 & 0\\
			2/3 & 2/3 & 1 & 0 & 0\\
			0 & 0 & 0 & 1 & \sqrt{2}/2\\
			0 & 0 & 0 & \sqrt{2}/2 & 1
		\end{pmatrix}.
	\end{equation}
	Here $E_W \equiv -\frac{2}{3}\log\frac{2}{3} - \frac{1}{3}\log\frac{1}{3} \approx 0.9183$. Similarly, by measuring such qubits in the mutually unbiased bases $\sigma_z$ and $\sigma_x$, one can check that the ideal uncertainty matrix is as follows:
	\begin{equation}
		\mathbb{U}^Q = \begin{pmatrix}
			1+E_W & U_W & U_W & 1+E_W & 1+E_W\\
			U_W & 1+E_W & U_W & 1+E_W & 1+E_W\\
			U_W & U_W & 1+E_W & 1+E_W & 1+E_W\\
			1+E_G & 1+E_G & 1+E_G & 1+E_G & E_G\\
			1+E_G & 1+E_G & 1+E_G & E_G & 1+E_G
		\end{pmatrix}.
	\end{equation}
	Since $E_G < 1$, we can still use Eq.\ (\ref{lower bound of distillable entanglement}) to detect the entanglement between qubits 4 and 5. However, the ideal entropic uncertainty of $W_{12}$ exceeds 1 despite its entanglement nature. This represents a limitation of employing the EUR for entanglement detection.
	
	Although the structure of the uncertainty matrix $\mathbb{U}^Q$ is less explicit compared to its counterparts, it remains useful for inferring the topology of quantum networks that involve entangled states beyond GHZ. In the above case of a $W$ state $|W\rangle$ and a generalized EPR state $|\text{EPR}^\prime\rangle$, the entropic uncertainty of the uncorrelated states is at least $1+E_G\approx 1.6009$. Therefore, a two-qubit uncertainty below the threshold of 1.6 certifies that the two qubits originate from the same source, even in the presence of noise. Recall that the uncertainty threshold for uncorrelated qubits in networks solely containing GHZ states is 2. In networks with a broader variety of entanglement sources, network users would need to incorporate knowledge about the types of entangled states to set this threshold for uncorrelated qubits, thereby enabling the reconstruction of network topology.

	Next, we perform variational optimization on the considered five-qubit network to compare the performance of different correlation matrices. Exploiting the identical simulation method in Sec.\ \ref{Comparison and simulation}, we plot the distances to ideal matrices and the heatmap of the optimized matrices in Fig.\ \hyperref[Distance_of_averaged_matrices_W]{10(a)} and\ \hyperref[Matrix_heatmap_W]{10(b)}, respectively. We find that the uncertainty-based and covariance-based matrices exhibit better numerical stability. In contrast, the mutual information of the generalized EPR state fails to reach its ideal value $E_G$ on average. 

\begin{figure}[t]
		\centering
		\includegraphics[scale=0.19]{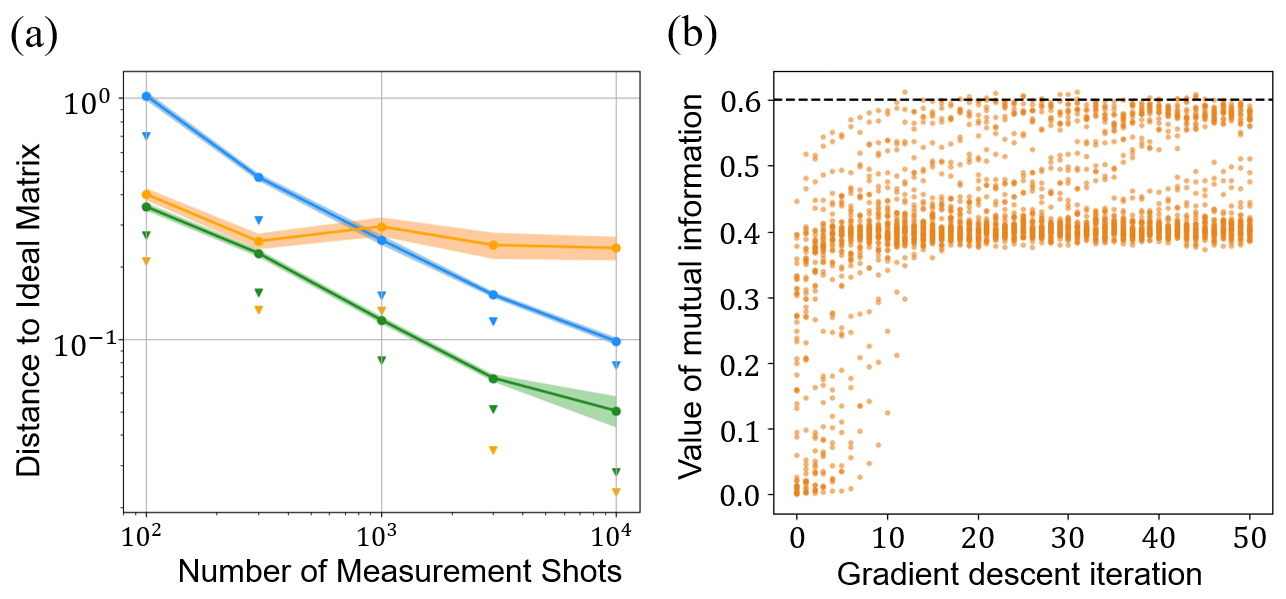}
		\caption{\label{Averaged_distance&two-qubit_optimization_W}(a) Averaged distances (circles with shaded standard error) and minimum distances (dashed triangles) of the optimized matrices across all trials as a function of the number of shots. We consider the five-qubit network consisting of a three-qubit $W$ state and a generalized EPR state. The distances of qubitwise uncertainty matrices (blue) and covariance matrices (green) properly decrease with increasing measurement shots. Yet for qubitwise characteristic matrices (orange), the effect of shot noise is dominated by the probabilistic failure of mutual information optimization. (b) Variational optimization of mutual information for the generalized EPR state. The scattered points denote values of optimized mutual information from 50 independent optimization trials with optimal step size 0.3. We take 10000 measurement shots to collect statistics. The black dashed line indicates the ideal value $E_G\approx 0.6009$, which is only achieved in a small fraction of trials.}
	\end{figure}

	Moreover, Fig.\ \hyperref[Averaged_distance&two-qubit_optimization_W]{11(a)} illustrates the averaged and minimum distances of the optimized matrices. The distances of uncertainty matrices in this example are relatively larger than those in the network of a 3-GHZ state together with an EPR state (see Fig.\ \ref{Averaged distances}). This degradation is attributed to the suboptimal value of the uncertainty of $W_{12}$, although it has little impact on the feasibility of network topology inference. Crucially, one can also see that the averaged distances between the qubitwise characteristic matrix and its ideal matrix do not decrease properly with increasing measurement shots, whereas the minimum distance does. As confirmed by a separate optimization of the bipartite state $|\text{EPR}^\prime\rangle$ in Fig.\ \hyperref[Averaged_distance&two-qubit_optimization_W]{11(b)}, the variational optimization of mutual information for generalized EPR states is also besieged by the local minima problem, similar to that for multipartite GHZ states. We observe that only a small fraction of optimization trials achieved the ideal value, while the majority converged around 0.4. Nevertheless, we acknowledge that for certain network topologies, the mutual information-based approach may exhibit particular computational advantages, such as in networks composed entirely of EPR pairs or $W$ states.
	
	\section*{Appendix B: Mitigating errors in a network}\label{Mitigating errors in a network}
	In this Appendix, we exploit several QEM techniques to suppress noise-induced errors in correlation matrices of quantum networks. Specifically, we employ the VD method introduced in Sec.\ \ref{Virtual distillation method} as well as the shadow distillation (SD) method \cite{seif_shadow_2023} to mitigate the errors in qubitwise matrices corresponding to the example five-qubit network in Fig.\ \ref{Example_five-qubit_network}. What is more, for another scenario where qubits in each node are measured collectively, SD is adopted to achieve more accurate evaluation of node-wise matrices, which further facilitates the application of Theorem \hyperref[Theorem 2]{2}.
	
	\begin{figure*}[ht]
		\begin{center}
			\includegraphics[scale=0.36]{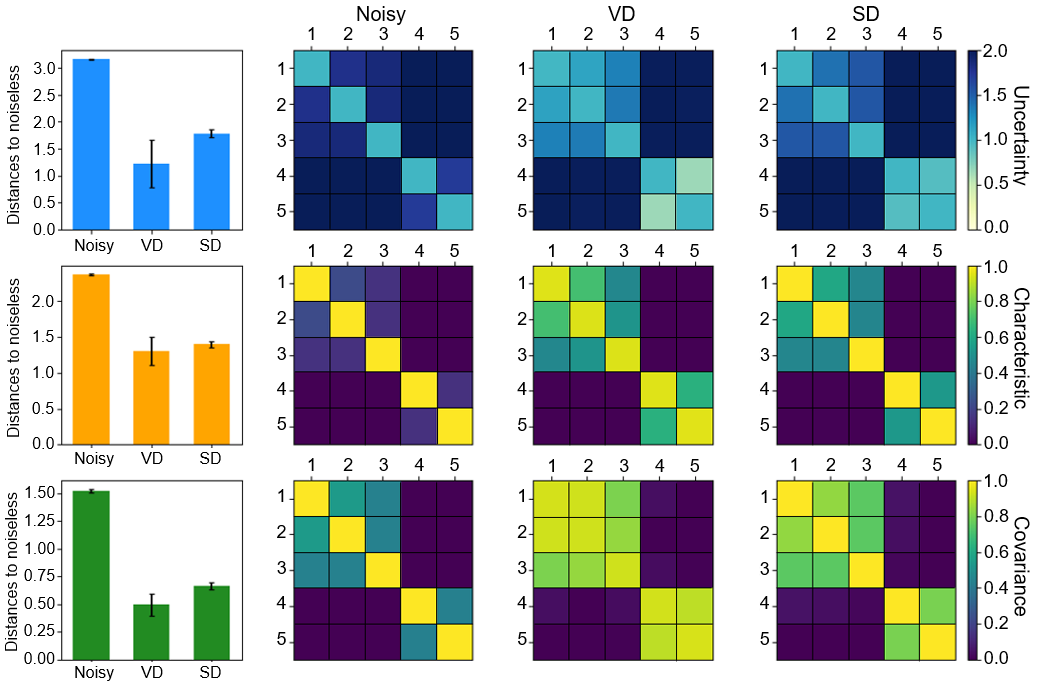}
			\caption{\label{Matrix_heatmap_VD&SD} Simulation of qubitwise correlation matrices for the noisy network consisting of a three-qubit GHZ state and an EPR state. From top to bottom are the results of qubitwise uncertainty matrix, qubitwise characteristic matrix, and covariance matrix. The three rightmost columns display correlation matrices obtained without QEM, with VD, and with SD, respectively, all averaged over ten independent simulations. The leftmost column plots the distance between these correlation matrices and the ideal noise-free matrix. Error bars represent standard deviations across ten simulation runs.}
		\end{center}
	\end{figure*}

	Recall that the previously introduced QEM methods typically use linear combinations of several measurement results to obtain the mitigated probabilities. However, this fact inevitably leads to negative mitigated probabilities when finite measurement shots are taken. In order to obtain valid probabilities in practical finite shot regime, we solve the following constrained optimization problem \cite{geller_rigorous_2020}
	\begin{equation}
		\begin{aligned}
			&\mathop{\arg\min}\limits_{\mathbb{P^\prime}}\| \mathbb{P^\prime}-\mathbb{P} \|^2_{2}\\
			\text{subject to  \ } & 0 \leq \mathbb{P^\prime}(x) \leq 1,\ \forall\ x \text{ and } \sum_{x} \mathbb{P^\prime}(x) = 1,
		\end{aligned}
	\end{equation}
	where $\mathbb{P}$ is the unphysical probability distribution that obtained through Eq.\ (\ref{error-mitigated joint probabilities}) or (\ref{error-mitigated VD probabilities}), $x$ denotes measurement outcome, e.g., $x = 00, 01, 10 ,11$ for two-qubit correlation, and $\|\cdot\|_2$ is the Euclidean distance between vectors. This optimization yields the physical probability distribution $\mathbb{P^\prime}$ that best approximates the unphysical mitigated one. $\mathbb{P^\prime}$ is then forwarded to calculations of entropic quantities and we take the calculation results as the final output of QEM method. Note that such a processing of probability distribution will hinder the variational optimization, which relies on gradient descent. Thus, we assume that a shared reference frame between all nodes has already been established in the following. This eliminates the need for variational optimization, and consequently the measurement bases are just $\sigma_z$ ($\sigma_z$ and $\sigma_x$) for mutual information (uncertainty) evaluation.

	Next, we briefly introduce the SD method \cite{seif_shadow_2023} applied in the quantum network scenario. Unlike the VD procedure, which requires coherent access to two copies of the network state, SD operates by measuring each qubit directly in a uniformly random Pauli basis. The random measurements output a string of outcomes $|b\rangle = |b_1, b_2, \dots, b_{N_Q}\rangle$. From these outcomes and the sampled measurement bases, a classical snapshot is reconstructed as
	\begin{equation}
		\hat{\rho}_{U,b} = \bigotimes_{k=1}^{N_Q}(3 U^\dagger_k |b_k\rangle\langle b_k| U_k - \mathbb{I}_2),
	\end{equation}
	where $b_k \in \lbrace 0, 1 \rbrace$ is the measurement outcome of the $k$th qubit and the single-qubit unitary $U_k \in \lbrace \mathbb{I}, H, HS\rbrace$ corresponds to Pauli measurements $\lbrace \sigma_z, \sigma_x, \sigma_y\rbrace$. Here, $H$ is the Hadamard gate and $S = |0\rangle\langle0| - \mathrm{i} |1\rangle\langle1|$. When the outcomes and the sampled unitaries are averaged, the classical snapshot serves as an unbiased estimator of the real density matrix, i.e, $\mathbb{E}_{U,b}(\hat{\rho}_{U,b}) = \rho$. Therefore, using Eqs.\ (\ref{error-mitigated expectation value}) and (\ref{Equality for VD}), the mitigated expectation value of an observable $O$ can be calculated as
	\begin{equation}
		\langle O \rangle_{\text{SD}} = \frac{\mathbb{E}_{U,b,U^\prime,b^\prime} [\operatorname{Tr}( S^{(2)} (O \hat{\rho}_{U,b}) \otimes \hat{\rho}_{U^\prime,b^\prime} )]}  { \mathbb{E}_{U,b,U^\prime,b^\prime} [\operatorname{Tr}( S^{(2)} \hat{\rho}_{U,b} \otimes \hat{\rho}_{U^\prime,b^\prime} ) ]},
		\label{SD estimator}
	\end{equation}
	where $S^{(2)} = \bigotimes_{k=1}^{N_Q} S_k^{(2)} $ denotes the swap operations between each pair of qubits across two copies. 

	In the context of network topology inference, our primary focus is on the correlations between subsystems of the network. Thus, we construct snapshots only for the relevant subsystems rather than for the global network state, thereby significantly reducing computational overhead. To obtain the mitigated probabilities for evaluating correlations, we can choose the observable $O$ in Eq.\ (\ref{SD estimator}) as either projectors onto the outcome strings, or Pauli observables $Z_i Z_j$ and subsequently use Eq.\ (\ref{error-mitigated VD probabilities}). Notably, when using projectors, the resulting probability distribution is inherently valid as ensured by the structure of Eq.\ (\ref{SD estimator}). However, when using Pauli observables, this validity is not guaranteed, necessitating a projection onto the closest valid probability distribution.

	Equipped with these methodologies, we now simulate the performance of VD and SD in the finite shot regime. We consider the five-qubit network in Fig.\ \ref{Example_five-qubit_network}, where qubits 1, 2, and 3 originate from three-partite GHZ source, whereas qubits 4 and 5 come from an EPR source. Independent depolarizing channels (see Eq.\ (\ref{Depolarizing Kraus operators})) with $\gamma = 0.2$ are applied to qubits 1, 2, and 4, while qubits 3 and 5 undergo depolarization with $\gamma = 0.1$. The heatmaps of ideal matrices in the noiseless scenario are depicted in the leftmost column of Fig.\ \hyperref[Matrix_heatmap]{6(a)}. 10 000 copies of the network state are utilized in simulation of each correlation matrix and we perform ten repetitions of the calculation. When applying VD, the copies are evenly divided for performing diagonalization gate $B_k^{(2)}$ (see Eq.\ (\ref{diagonalization gate B})) and $D_k^{(2)}$ (see Eq.\ (\ref{diagonalization gate D})). A better division might reduce the variance of results. The averaged correlation matrices with and without the application of QEM techniques, along with their corresponding Euclidean distances to the noiseless matrices, are illustrated in Fig.\ \ref{Matrix_heatmap_VD&SD}. All mitigated matrices show closer proximity to their noiseless counterparts. Particularly, the implementations of both methods enable the uncertainty between qubits 4 and 5 to drop below the entanglement detection threshold of 1. These improvements demonstrate successful error suppression and more robust identification of EPR source origination under strong physical noise.

	\begin{figure}[!b]
		\centering
		\includegraphics[scale=0.27]{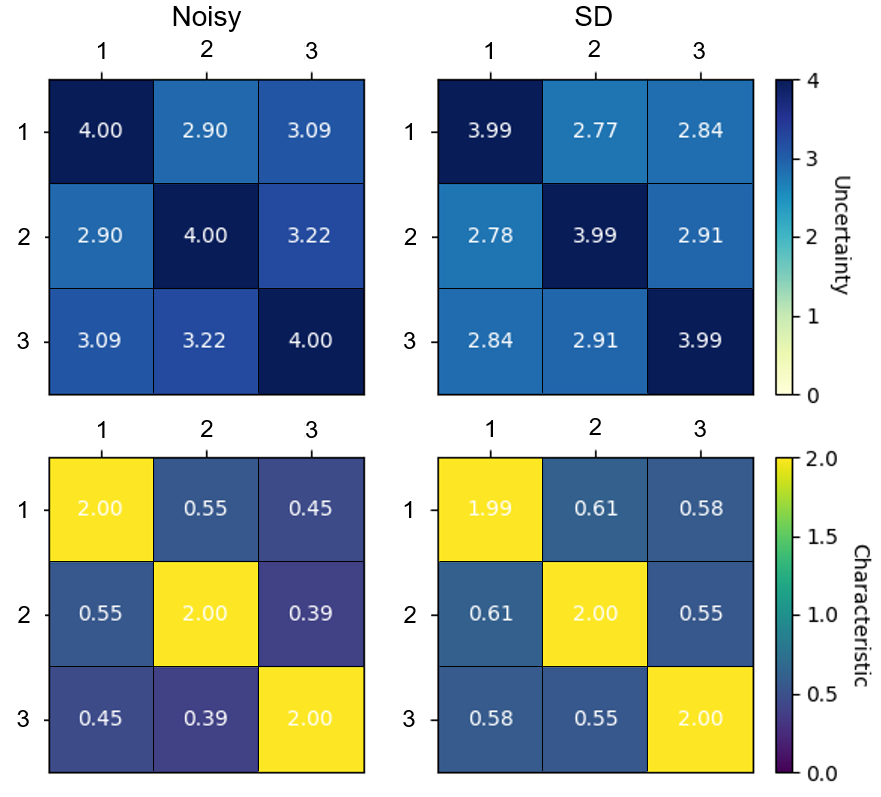}
		\caption{\label{Matrix_heatmap_3EPR_SD} Simulation of node-wise correlation matrices for the noisy network in Fig.\ \hyperref[Example_networks]{2(a)}. The calculation of each matrix consumes 10 000 copies of the network state. The annotated numbers indicate the values of matrix entries averaged over ten runs.}
	\end{figure}

	Furthermore, we extend the application of SD to the scenario of collective measurements. As a paradigmatic demonstration, we consider the triangle network in Fig.\ \hyperref[Example_networks]{2(a)} with three nodes interconnected pairwise via three EPR sources. The node-wise uncertainty matrix and characteristic matrix of this network in noiseless scenario are
	\begin{equation}
		\mathbb{U} = \begin{pmatrix}
			4 & 2 & 2\\
			2 & 4 & 2\\
			2 & 2 & 4
		\end{pmatrix},\ M=\begin{pmatrix}
			2 & 1 & 1\\
			1 & 2 & 1\\
			1 & 1 & 2
		\end{pmatrix}.
	 \end{equation}
	To simulate realistic noise conditions, we impose distinct depolarization channels on the qubits transmitted to nodes $C_1, C_2$, and $C_3$, with $\gamma$= 0.05, 0.10, and 0.15, respectively. As evidenced by the obtained correlation matrices in Fig.\ \ref{Matrix_heatmap_3EPR_SD}, the application of SD effectively reduces the deviations in uncertainties and mutual information. 

	Recall that in a noiseless quantum network, the number of shared EPR sources between two nodes can be counted through the relation $N_{A,ij}^{\text{EPR}} = H_{u}(C_i) - H_{u}(C_i | C_j) - I(C_i, C_j)$ (see Eq.\ (\ref{Counting EPR})). In practical noisy circumstances, however, noise inherently smears out the measured correlations by increasing uncertainty and decreasing mutual information. Besides, statistical fluctuations or physical noise such as amplitude damping will reduce Shannon entropies. To compensate for the noise-induced distortions, we suggest applying discontinuous functions to realistic correlations. Specifically, ceiling functions are applied to quantities that decrease due to noise, while a floor function to the two-qubit uncertainty. This leads to the modified expression:
	 \begin{equation}
		N_{A,ij}^{\text{EPR}} = \lceil H_{u}(C_i) \rceil - \lfloor H_{u}(C_i | C_j) \rfloor - \lceil I(C_i, C_j) \rceil
	\end{equation}
	This adjustment moderately enhances the robustness of Theorem \hyperref[Theorem 2]{2}. However, as one can see from the noisy correlation matrices in Fig.\ \ref{Matrix_heatmap_3EPR_SD}, the efficacy of this approach diminishes when the number of involved qubits is large or when noise levels are high. Of note, when integrated with SD, the modified relation correctly counts the number of EPR pairs for all node pairs, exhibiting improved noise tolerance without additional entanglement resources. This makes SD a promising candidate for near-term quantum network implementations where full error correction is constrained.

	\bibliography{References}
\end{document}